\documentclass[11pt,a4paper]{article}

\usepackage{jheppub} 
\usepackage{amsmath}
\usepackage{amsfonts}
\usepackage{amssymb}
\usepackage{graphicx} 
\usepackage{epsfig}
\usepackage{color}
\usepackage{amssymb}
\usepackage{bbold}
\usepackage{hhline}
\usepackage{tabu}

\usepackage{epstopdf}
\usepackage{placeins}
\newcommand{\dd}{\mathrm{d}}

\def\beq{\begin{equation}}
\def\eeq{\end{equation}}
\def\baq{\begin{eqnarray}}
\def\eaq{\end{eqnarray}}
\newcommand{\ee}[1]{\begin{equation}#1\end{equation}}
\newcommand{\ea}[1]{\begin{align}#1\end{align}}

\newcommand{\da}{\ensuremath{\dot{a}}}
\newcommand{\dda}{\ensuremath{\ddot{a}}}

\providecommand{\f}[2]{\frac{{#1}}{{#2}}}

\title{The 1-loop effective potential for the Standard Model in curved spacetime}

\author[a]{Tommi Markkanen}
\author[b]{Sami Nurmi}
\author[c]{Arttu Rajantie}
\author[d]{Stephen Stopyra}
\affiliation[a,c,d]{Department of Physics, Imperial College London, SW7 2AZ, UK}
\affiliation[b]{Department of Physics, University of Jyva\"skyl\"a, P.O. Box 35, FI-40014 University of Jyva\"skyl\"a, Finland}

\abstract{
The renormalisation group improved Standard Model effective potential in {an arbitrary} curved spacetime is computed to one loop order in perturbation theory. 
The loop corrections are computed in the ultraviolet limit, which makes them independent of the choice of the vacuum state and allows the derivation of the complete set of $\beta$-functions. The potential depends on the spacetime curvature through the direct non-minimal Higgs-curvature coupling, curvature contributions to the loop diagrams, and through the curvature dependence of the renormalisation scale. Together, these lead to significant curvature dependence, which needs to be taken into account in cosmological applications, which is demonstrated with the example of vacuum stability in de Sitter space.}

\emailAdd{t.markkanen@imperial.ac.uk}
\emailAdd{sami.t.nurmi@jyu.fi}
\emailAdd{a.rajantie@imperial.ac.uk}
\emailAdd{stephen.stopyra09@imperial.ac.uk}
\begin{document} 
\begin{flushleft}
	\hfill		  IMPERIAL/TP/2018/TM/02
\end{flushleft}
\maketitle
\section{Introduction}
Ever since the seminal work \cite{Coleman:1973jx} the quantum corrected or effective potential has been amongst the principal tools of quantum field theory. The effective potential 
in curved spacetime can have a number of important cosmological impacts. A key example is the analysis of vacuum stability in the early universe. The Standard Model (SM) of particle physics predicts a metastable electroweak vacuum  \cite{Chigusa:2017dux,Chigusa:2018uuj,Andreassen:2017rzq,Bednyakov:2015sca,Degrassi:2012ry,Buttazzo:2013uya,Bezrukov:2012sa,Espinosa:1995se, Isidori:2001bm,Ellis:2009tp}. Its survival over inflation and reheating is a non-trivial consistency requirement both for the SM and its extensions \cite{vacstab,vacstab1,vacstab2,vacstab3,vacstab4,vacstab5,vacstab6,vacstab7,vacstab8,vacstab9, vacstab10,vacstab11,vacstab12,vacstab13,vacstab14,vacstab15,vacstab16,vacstab17,East:2016anr, vacstab19,vacstab20,vacstab21,Moss:2015gua,Figueroa:2017slm, Ema:2017ckf,Branchina:2016bws,Bentivegna:2017qry,Branchina:2013jra}. The stability conditions crucially depend on curved spacetime  contributions \cite{vacstab,vacstab7, Herranen:2015ima} in the effective potential which affect the behaviour of energetically subdominant spectator fields such as the SM Higgs.  In SM extensions, radiatively generated curvature couplings can also produce primordial dark matter \cite{Markkanen:2015xuw}. Smallness of curvature induced mass terms is also a key condition required in the curvaton scenario \cite{curvaton} where massless spectator scalars source the primordial perturbation. 

From a more fundamental point of view, in a quantum field theory setting the inclusion of gravity in the form of background curvature leads to interesting an important modifications: the renormalization group (RG) running scale generically is influenced by the curvature leading to \textit{curvature induced running}. The importance of this effect was first discovered in \cite{vacstab7} and has since been shown to give rise to significant consequences in various set-ups \cite{Herranen:2015ima,Herranen:2016xsy,Markkanen:2014poa,East:2016anr}. Another crucial feature resulting from background gravity is the generation of new gravity-dependent operators, most famously of the non-minimal coupling between scalar fields and the scalar curvature of space, as already discussed in \cite{Chernikov:1968zm,Callan:1970ze,Tagirov:1972vv}. These profound features are not visible in an approximation that neglects the curvature of the background.

Making generic statements about the behaviour of a spectator field in curved spacetime is unfortunately hindered by the calculational complexity of the problem: deriving the complete effective potential in an arbitrary curved spacetime is in general quite involved and obtaining explicit results requires one to specify the set-up, including making a choice for the background and the quantum state of interest. For examples of such calculations, see  \cite{Calzetta:1986ey,Boyanovsky:1993xf,Boyanovsky:2005sh,Miao:2006pn,Bilandzic:2007nb,Janssen:2009pb,Serreau:2011fu,Herranen:2013raa}. There are however some aspects that are universal. According to standard field theory principles, the ultraviolet (UV) behaviour of a theory must be state independent in order to have unique divergent parts in the counter terms that are required for rendering the theory finite. Furthermore, since techniques are available with which to extract the UV contribution to the effective potential in a general curved spacetime, deriving the complete set of operators generated by the quantum corrections as well as investigating the RG running of constants can be performed without choosing a specific form of the background metric or the quantum state \cite{Birrell:1982ix}.

In this work we calculate the UV contribution to the effective potential for the SM Higgs, in an arbitrary curved spacetime including all degrees of freedom contained in the SM to 1-loop order. We furthermore derive the complete set of $\beta$-functions with which we perform renormalization group improvement of the result. We will throughout work in the approximation where the SM Higgs is a subdominant spectator while neglecting the metric fluctuations, which has been shown to be a very good approximation \cite{Markkanen:2017dlc}. Recently similar calculations, primarily in the context of the SM vacuum instability during inflation, have been performed in \cite{vacstab7,vacstab15,Bounakis:2017fkv,Kohri:2016qqv} and see \cite{Hu:1984js,Elizalde:1993ee,Herranen:2016xsy,Kirsten:1993jn,Markkanen:2012rh,George:2012xs,Buchbinder:1986yh,Elizalde:1993ew,Elizalde:1994im,Toms:1982af,Toms:1983qr,Odintsov:1993rt,Elizalde:1993qh,Odintsov:1990mt,Ford:1981xj} for related earlier studies. {We however emphasize that the current work is the first one to present the complete result i.e. it includes all degrees of freedom of the SM along with all operators generated by quantum corrections in curved space.} Our calculation is based on the well-known Heat Kernel technique \cite{Schwinger:1951nm,DeWitt:1965jb,Seeley:1967ea,Gilkey:1975iq,Minakshisundaram:1949xg,Hadamard,Avramidi:2000bm}, which is essentially a gradient expansion, and we will in particular make use of the resummed form presented in \cite{Parker:1984dj,Jack:1985mw}.  

We will also implement our result in the specific case of the de Sitter background and revisit the analysis of electroweak vacuum stability during inflation. Requiring that the electroweak vacuum survives inflation, we compute the lower bound for the non-minimal coupling as function of the SM Higgs and top quark masses. As a new result, we show that negative values of the non-minimal coupling are tightly constrained from below even if the inflationary scale is well below the instability scale $\mu_{\rm inst}$ where 
$\lambda(\mu_{\rm inst}) = 0$. This sets a non-trivial lower bound on the non-minimal coupling even for low top mass values for which $\mu_{\rm inst}$ is larger than the maximal inflationary scale allowed by the non-detection of primordial gravitational waves. 

Our sign conventions for the metric and curvature tensors are $(-,-,-)$ in the classification of \cite{Misner:1974qy}.

\section{Effective potential for a self-interacting scalar field}
\label{sec:effp}
The derivation of the effective potential for a scalar field in an arbitrary curved spacetime for theories containing scalar, fermion and gauge fields will be addressed in section \ref{eq:appa} and implemented for the full SM in de Sitter space in section \ref{sec:SM}. But first for illustrative purposes we will show the necessary steps by using the self-interacting scalar field as a toy model. Although simple, this model will exhibit all the qualitative features that arise in more complicated theories when background curvature is not neglected in the derivation of the effective potential. We will also discuss renormalization group (RG) improvement in curved spacetime in this context. A point worth emphasizing is that we are only interested in behaviour at the very high ultraviolet (UV) limit. This stems from the fact only the UV is relevant when discussing the radiative generation of operators not present at tree-level and relatedly determining the RG running and the $\beta$-functions. For this reason we can make use a large momentum approximation throughout, which will simplify the derivation considerably. 

The action for some generic massive, non-minimally coupled and self-interacting scalar field $\chi$ reads
\ee{S_m=\int d^4x\,\sqrt{-g}\bigg[\f{1}{2}\partial_\mu\chi_0\partial^\mu\chi_0-\f{1}{2}m^2_0\chi_0^2-\f{\xi_0}{2}R\chi_0^2-\f{\lambda_0}{4}\chi_0^4\bigg]\,,\label{eq:actS}}
where $R$ is the scalar curvature. The subscripts "0" indicate bare or unrenormalized parameters. In curved spacetime proper renormalization requires one also to introduce a purely gravitational part to the action as such operators are radiatively generated \cite{Birrell:1982ix,ParkerToms}. As we will show, the running of these at tree-level purely gravitational operators will turn out to be important for the effective potential. Specifically, the gravitational action reads 
\ee{S_g=-\int d^4x\,\sqrt{-g}\bigg[V_{\Lambda,0}-\kappa_0 R+\alpha_{1,0} R^2+\alpha_{2,0} R_{\mu\nu}R^{\mu\nu}+\alpha_{3,0} R_{\mu\nu\delta\eta}R^{\mu\nu\delta\eta}\bigg]\label{eq:treecurve}\,,}
where\footnote{These lead to the traditional parametrization of the Einstein equation with
 \ee{R_{\mu\nu}-\f{1}{2}Rg_{\mu\nu}+g_{\mu\nu}\Lambda_0=-8\pi G_0 T_{\mu\nu}\,;\qquad \f{2}{\sqrt{-g}}\f{\delta S_m}{\delta g^{\mu\nu}}=T_{\mu\nu}\,.}
}  $\kappa_0 = (16\pi G_0)^{-1}$ and $V_{\Lambda,0}=(8\pi G_0)^{-1}\Lambda_0$.
 Since we assume an unbounded space the terms $\Box\chi^2$ and $\Box R$ are not present in the action as they may be removed by partial integration.

For a scalar field in the 1-loop approximation the effective potential can be studied without the need of more sophisticated approaches, namely the Heat Kernel technology~\cite{Schwinger:1951nm,DeWitt:1965jb,Seeley:1967ea,Gilkey:1975iq,Minakshisundaram:1949xg,Hadamard,Avramidi:2000bm}. For more complicated theories however, the Heat Kernel approach does prove to be extremely convenient as will become apparent in the following two sections, but for a model containing a single scalar field the derivation can be completed by simply making use of the equations of motion. 

The derivation we are about to present is somewhat simpler than the traditional one found in the seminal work \cite{Coleman:1973jx} and standard textbooks \cite{Peskin:1995ev,Cheng:1985bj}, mainly because it does not rely on an infinite summation of one-particle-irreducible Feynman diagrams and hence the often non-trivial concept of symmetry factors never comes up. But more importantly for our purposes, the derivation can very easily be generalized to curved spaces.

In order to derive the quantum corrected or effective equations of motion we shift the quantized field as \ee{{\hat{\chi}_0}\rightarrow\langle\hat{\chi}_0\rangle+\hat{\chi}_0\equiv \chi_0+\hat{\chi}_0\,,\label{eq:fluc}} to 1-loop order the equation of motion for the mean field $\chi$ and the fluctuation $\hat{\chi}$ can be derived by first expanding the action to quadratic order
\ea{\label{eq:expS}S_m=&-\f{1}{2}\int d ^4x\sqrt{-g}~ \bigg[-\partial_\mu\chi_0\partial^\mu\chi_0+m^2_0\chi^2_0 +\xi_0 R\chi_0^2+2\f{\lambda_0}{4}\chi_0^4\bigg]\nonumber \\ &-\f{1}{2}\int d ^nx\sqrt{-g}~\hat{\chi}_0\bigg[\Box+\xi_0 R+M^2(\chi_0)\bigg]\hat{\chi}_0+\cdots\,,}
where we have defined the flat space effective mass
\ee{M^2(\chi)=m^2_0+3{\lambda_0}\chi_0^2\,.\label{eq:effM2}}
The above leads to two coupled equations, one for the mean field and one for the fluctuation
\ea{&\bigg[\Box+m_0^2 +\xi_0 R+\lambda_0\chi_0^2\bigg]\chi_0+3{\lambda_0}\chi_0\langle\hat{\chi}_0^2\rangle=0\,,\label{eq:eom2}\\ &\bigg[\Box +\xi_0 R+M^2(\chi_0)\bigg]\hat{\chi_0}=0,\label{eq:eom3}
}
Note that to this order of truncation the diagrams containing an odd number of external legs drop out.

The counter terms 
are obtained by defining the renormalized field with the wave function renormalization factor $Z$ \cite{Peskin:1995ev}
\ee{\chi_0=\sqrt{Z}\chi\label{eq:Z0}\,,}
and similarly setting $Z=1+\delta Z$, $Zm_0^2=m^2+\delta m^2$ and $Z^2\lambda_0= \lambda+\delta\lambda$.

For a constant mean field $\chi$ the renormalized quantum corrected equation of motion (\ref{eq:eom2}) reduces to finding the minimum of the effective potential, which to 1-loop order can be written as
\ee{V_{\rm eff}'(\chi)=\bigg[\xi R+m^2 +\lambda\chi^2\bigg]\chi +3{\lambda}\chi\langle\hat{\chi}^2\rangle-\delta V'(\chi)=0\,,\label{eq:effM}} where $\delta V(\chi)$ contains the counter terms for which from now on we use the unifying notation $\delta c_i$.

The effective potential straightforwardly follows from integration
\ee{
V_{\rm eff}(\chi)\equiv V^{(0)}(\chi)+V^{(1)}(\chi)+\cdots =\int^\chi V_{\rm eff}'(\tilde{\chi})\,d\tilde{\chi}\,,\label{eq:effV}}
where the superscripts $"(0)"$ and $"(1)"$ denote the tree-level and 1-loop pieces, respectively.

For a scalar field, to 1-loop order finding a solution in the UV approximation for the quantum field in terms of modes is relatively simple even when the curvature of the background is included in the discussion. For completeness however we first present the derivation in flat space.

\subsection{1-loop in flat space}
As usual, in flat space the solutions to (\ref{eq:eom3}) to 1-loop order can be expressed as a mode expansion
\ee{\hat{\chi}=\int \f{d^{3}\mathbf{k}}{\sqrt{(2\pi)^3}}\, e^{i\mathbf{k\cdot\mathbf{x}}}\left[\hat{a}_\mathbf{k}^{\phantom{\dagger}}f^{\phantom{\dagger}}_{k}(t)+\hat{a}_{-\mathbf{k}}^\dagger f^*_k(t)\right]\,\label{eq:adsol3}\,;\qquad f^{\phantom{\dagger}}_{k}(t)=\f{e^{-i\omega t}}{\sqrt{2\omega}}\,;\qquad \omega^2\equiv\mathbf{k}^2+M^2(\chi)\,,}
where $[\hat{a}_{\mathbf{k}}^{\phantom{\dagger}},\hat{a}_{\mathbf{k}'}^\dagger]=\delta^{(3)}(\mathbf{k}-\mathbf{k}'),~~[\hat{a}_{\mathbf{k}}^{\phantom{\dagger}},\hat{a}_{\mathbf{k}'}^{\phantom{\dagger}}]=[\hat{a}_{\mathbf{k}}^{{\dagger}},\hat{a}_{\mathbf{k}'}^\dagger]=0$ and
$\mathbf{k}$ is the momentum with $k\equiv|\mathbf{k}|$. The effective mass $M^2(\chi)$ is found from (\ref{eq:effM2}).
It is then trivial to use the mode solution and write
\ee{V_{\rm eff}'(\chi)=m^2\chi+{\lambda}\chi^3+3{\lambda}\chi\int\f{d^3\mathbf{k}}{2(2\pi)^3}\f{1}{\sqrt{\mathbf{k}^2+M^2(\chi)}}-\delta V'(\chi)\,, \label{eq:Veffsca}}
which by performing the integral over $\chi$ as in (\ref{eq:effV}) and using the standard formulae for dimensional regularization \cite{Peskin:1995ev}
gives the 1-loop effective potential
\ea{V_{\rm eff}(\chi)
&=\f{1}{2}m^2{\chi}^2 +\f{\lambda}{4}{\chi}^4+\f{M^4(\chi)}{64\pi^2}\bigg[\log\left(\f{M^2(\chi)}{\mu^2}\right)-\f{3}{2}+\Big\{-\f{2}{\epsilon}-\log(4\pi)+\gamma_e\Big\}+\mathcal{O}(\epsilon)\bigg]\nonumber \\ &-\bigg[\delta V_\Lambda+\f{1}{2}\delta m^2{\chi}^2 +\f{\delta\lambda}{4}{\chi}^4\bigg]
\label{eq:Veffsca2}\,,}
where the divergences are expressed in terms of $n=4-\epsilon$ and we have introduced the usual renormalization scale $\mu$. In the $\overline{\rm MS}$ subtraction scheme, which we will from now on use throughout, the divergent pole at $n\rightarrow4$, the $\log (4\pi)$ and the Euler constant in the wavy brackets would be removed by a proper choice of the renormalization counter terms, $\delta V_\Lambda,\delta m^2$ and ${\delta\lambda}$. Note that even in flat space a divergence $\propto m^4$ is generated and strictly speaking the cosmological constant counter term $\delta V_\Lambda$ introduced by the gravitational action (\ref{eq:treecurve}) is required.

\subsection{1-loop in curved spacetime}
\label{eq:1cS}
In curved spacetime we can define a properly normalized ansatz for the modes by first restricting our background to a homogeneous and isotropic one described via the Friedmann--Lema\^itre--Robertson--Walker (FLRW) metric given in cosmic time as \ee{ds^2=dt^2-a^2d\mathbf{x}^2\,,}
then rescaling the field as in the previous section and finally writing
\ee{\hat{\chi}=\int\f{d^{3}\mathbf{k}}{\sqrt{(2\pi a)^3}}\, e^{i\mathbf{k\cdot\mathbf{x}}}\left[\hat{a}_\mathbf{k}^{\phantom{\dagger}}f^{\phantom{\dagger}}_{k}(t)+\hat{a}_{-\mathbf{k}}^\dagger f^*_k(t)\right]\,\,;\qquad {f}_{k}(t)=\f{e^{-i\int^{t}Wdt'}}{\sqrt{2W}}\label{eq:ans}\,,}
which after inserting into the equation of motion for the fluctuation (\ref{eq:eom3}) gives a relation for $W$
\ee{\label{eq:W}W^2=
\f{k^2}{a^2}+M^2(\chi)+\f{\dda}{a}\f{3}{2}(4\xi-1)+\bigg(\f{\da}{a}\bigg)^2\f{3}{4}\big(8\xi-1\big)+\f{3\dot{W}^2}{4W^2}-\f{\ddot{W}}{2W}\,.}
Importantly, in practice the ansatz (\ref{eq:ans}) provides useful solutions only as a high momentum expansion. This is also the reason why it and the results that follow resemble very much the flat space results of the previous subsection: when probing the very high UV the global structure of spacetime is not visible as locally any smoothly curved manifold is nearly flat. 

We will solve for $W$ from (\ref{eq:W}) iteratively as an expansion in terms of large $k/a$. The first few orders may be written as
\ee{W=\sqrt{ \f{k^2}{a^2}+M^2(\chi)+\left(\xi- 1/6\right)R+\mathcal{O}(k/a)^{-2}}\,,}
which contain all terms leading to divergences in four dimensions and where $R$ is again the scalar curvature. It is now straightforward to calculate the 1-loop contribution to the variance, which can again be calculated with standard dimensional regularization 
\ea{\langle \hat{\chi}^2\rangle&=\f{\mu^\epsilon}{2}\int \f{d^{n-1} k}{(2\pi a)^{n-1}}\f{1}{\sqrt{(k/a)^2+M^2(\chi)+\left(\xi- 1/6\right)R}}\nonumber \\&=\f{M^2(\chi) +\left(\xi- 1/6\right)R}{16\pi^2}\bigg[\log\bigg(\f{M^2(\chi) +\left(\xi- 1/6\right)R}{\mu^2}\bigg)-1-\f{2}{\epsilon}-\log(4\pi)+\gamma_e\bigg] \,.\label{eq:varia}}
Like in the previous section by using (\ref{eq:effM}) and (\ref{eq:effV}) and choosing the appropriate counter terms we can write the 1-loop correction to the renormalized curved spacetime effective potential in a form very similar to the flat space result in (\ref{eq:Veffsca2})
\ea{V^{(1)}(\chi)
=\f{\left(M^2(\chi)+\left(\xi- 1/6\right)R\right)^2}{64\pi^2}\bigg[\log\left(\f{|M^2(\chi)+\left(\xi- 1/6\right)R|}{\mu^2}\right)-\f{3}{2}\bigg]+\mathcal{O}(R^2)\label{eq:curve}\,.}
A few comments are now in order. The notation $\mathcal{O}(R^2)$ indicates an inherent ambiguity in the derivation in terms of operators that are purely gravitational at tree-level: any contribution $\propto$ $R^2\log$, $R_{\mu\nu}R^{\mu\nu}\log$ or $R_{\mu\nu\delta\eta}R^{\mu\nu\delta\eta}\log$ results in a finite contribution for $V_{\rm eff}'(\chi)$ and will thus be invisible to a derivation including only the divergent terms in the effective equation of motion. We have also neglected any possible imaginary part of the effective
potential by using an absolute value in the logarithm. It is well-known from flat space that integration over the infrared modes may give rise to a complex result for the effective potential which is usually taken to indicate a finite lifetime of the state \cite{Weinberg:1987vp}, however this effect is not correctly represented in an approach that is based on an UV expansion. Furthermore, we have left in the same non-logarithmic finite pieces that are generated in the flat space $\overline{\rm MS}$ prescription (cf. eg. (\ref{eq:Veffsca2})).

As (\ref{eq:curve}) clearly shows, the $\mathcal{O}(R^2)$-type terms couple to the scalar field and are thus relevant for the effective potential. Next we will briefly present their derivation for the self-interacting scalar field model.

\subsubsection{via Heat Kernel techniques}
The derivation of the previous section via an UV expansion for the 1-loop approximation is to illustrate the modifications that arise when background curvature is not neglected. For deriving the effective potential for a theory including also fermions and gauge fields it becomes apparent that more sophisticated (and formal) technology is needed, namely the Heat Kernel techniques to be discussed in section \ref{eq:appa}. This is also useful for obtaining all the $\mathcal{O}(R^2)$ terms in (\ref{eq:curve}). 

Functional determinants 
are widely used in quantum field theory in flat space and we refer the reader to \cite{Peskin:1995ev} for more discussion for their use in traditional particle physics.
In this regard, we can express the 1-loop quantum correction from (\ref{eq:expS}) via a 'tracelog'
\ee{\int d^4x\,\sqrt{-g}\,V^{(1)}(\chi)= -\f{i}{2}{\rm Tr}\log \Big[\Box+M^2(\chi)+\xi R\chi^2\Big]\,,\label{eq:trlog1}}
as is well-known. This approach can also be generalized to the case of a curved spacetime. The detailed derivation and formulae may be found in section \ref{eq:appa}, but here we will simply apply the results of section \ref{eq:appa}, specifically subsection (\ref{sec:sca}) to (\ref{eq:trlog1}) in order to write 
\ea{
V_{\rm eff}(\chi)&=\f{1}{2} m^2\chi^2+\f{\xi}{2}R\chi^2+\f{\lambda}{4}\chi^4+V_\Lambda-\kappa R+\alpha_1 R^2+\alpha_2 R_{\mu\nu}R^{\mu\nu}+\alpha_3 R_{\mu\nu\delta\eta}R^{\mu\nu\delta\eta}\nonumber \\&+
\f{\left(M^2(\chi)+\left(\xi- 1/6\right)R\right)^2}{64\pi^2}\bigg[\log\left(\f{|M^2(\chi)+\left(\xi- 1/6\right)R|}{\mu^2}\right)-\f{3}{2}\bigg]\nonumber\\ &+\f{\f{1}{90}\left(R_{\mu\nu\delta\eta}R^{\mu\nu\delta\eta}-R_{\mu\nu}R^{\mu\nu}\right)}{64\pi^2}\bigg[\log\left(\f{|M^2(\chi)+\left(\xi- 1/6\right)R|}{\mu^2}\right)\bigg]
\label{eq:curve2}\,.}
The operators that are generated via radiative corrections in curved spacetime can be seen from the 1-loop correction in (\ref{eq:curve2}), which is why they needed to be present already at tree-level in (\ref{eq:actS}--\ref{eq:treecurve}) and are a part of the complete $V_{\rm eff}(\chi)$. Furthermore, as all operators couple to the renormalization scale $\mu$ they cannot be made to vanish for all scales which is felt in the dynamics of the scalar field due to the $\chi$-dependence of the logarithms in (\ref{eq:curve2}).

\subsection{RG improvement in the presence of curvature}
\label{sec:RGC}
Here we perform the RG analysis of the self-interacting scalar field model (\ref{eq:actS}--\ref{eq:treecurve}) and discuss RG improvement in curved space. Early work on RG improving the effective potential in flat space may be found in  \cite{Kastening:1991gv,Ford:1992mv,Bando:1992np,Bando:1992wy}. Studies in curved spacetime include \cite{Hu:1984js,Elizalde:1993ew,Buchbinder:1986yh,Elizalde:1993ee,Elizalde:1994im,Kirsten:1993jn,Toms:1982af,Toms:1983qr,Odintsov:1993rt,Elizalde:1993qh,Odintsov:1990mt,Ford:1981xj}, see also the textbook \cite{Buchbinder:1992rb}.

The Callan-Symanzik equation is first and foremost an expression of renormalization scale invariance: in principle the renormalization scale $\mu$ is an arbitrary choice and physical quantities should not depend on it. For the effective potential this translates as demanding \ee{\f{d V_{\rm eff}(\chi) }{d\mu}=0\,,\label{eq:csi}}
where we emphasize due to the coupling between $\chi$ and all the gravitational operators in (\ref{eq:treecurve}) the above includes all operators in the original action as visible in (\ref{eq:curve2}).

The requirement in (\ref{eq:csi}) leads to the well-known Callan-Symanzik equations for the effective potential in terms of the $\beta$-functions and the anomalous dimension $\gamma$ \cite{Ford:1992mv} 
\ee{\bigg\{\mu\f{\partial}{\partial\mu}+\beta_{c_i}\f{\partial}{\partial{c_i}}-\gamma \chi\f{\partial}{\partial\chi}\bigg\}V_{\rm eff}(\chi)=0\,, \qquad\beta_{c_i}\equiv \mu\f{\partial c_i}{\partial\mu}\,,\quad\gamma\equiv \mu\f{\partial \log \sqrt{Z}}{\partial\mu} \, ,\label{eq:CS}}
where the $c_i$ stands for all the parameters of the action with summation over the repeated index $i$ assumed. $Z$ is the wave function renormalization introduced in (\ref{eq:Z0}) defining the renormalized field. It should be kept in mind that the wave function renormalization has a dependence on the renormalization scale $\mu$ and so does then the renormalized field, $\chi(\mu)={Z(\mu)}^{-1/2}\chi_0$.

To 1-loop order, the above can be written as
\ee{\bigg\{\beta_{c_i}\f{\partial}{\partial{c_i}}-\gamma \chi\f{\partial}{\partial\chi}\bigg\}V^{(0)}(\chi)=-\mu\f{\partial}{\partial\mu}V^{(1)}(\chi)\, .\label{eq:CS1}}
In general one needs the anomalous dimension $\gamma$ as an input before the equation may be solved, which requires its determination by means other than the effective potential, for example from a direct evaluation of the 2-point function. This poses no additional complications in curved spacetime over the usual flat space case, since the $\gamma$ in a curved spacetime derivation must be identical to the flat space result and can be taken from standard literature. This is because the gravitational couplings such a $\xi$ and the $\alpha$'s in eq.~(\ref{eq:treecurve}) must only couple to operators that vanish in the flat space limit. If this were not the case and $\gamma$ did contain e.g. a contribution from $\alpha_1$ it would indicate that the size of a gravitational operator could also affect the running of all parameters in flat space, which is not tenable from a purely physical point of view.

Given $\gamma$, one may solve for the $\beta$-functions and the running constants and finally use them to improve the limit of applicability of the effective potential, with the end result in principle should be independent of $\mu$. As one can show however, this is not the case for any perturbative result, but rather there is always some residual $\mu$-dependence left, which is an artefact of our inability to solve the effective potential exactly. In \cite{Ford:1992mv} it was first proposed that in order to minimize the error from the neglected higher order terms the scale $\mu$ can be chosen such that the logarithms in the loop correction remain small. Choosing a particular form for $\mu$ is allowed since the complete result must be independent of $\mu$ as demanded by (\ref{eq:CS}). 

As an example we first discuss RG improvement in the 1-loop approximation in flat space for the simple scalar field model in (\ref{eq:actS}). The $\beta$-functions to 1-loop order are easy to derive by using (\ref{eq:Veffsca2}) and (\ref{eq:CS}), and noting that to 1-loop order the anomalous dimension vanishes, $\gamma=\mathcal{O}(\lambda^2)$, for scalar field with only a quartic interaction term,  which results in
\ee{\beta_\lambda=\f{9\lambda^2}{8\pi^2}\,;\qquad\beta_{m^2}=\f{3m^2\lambda}{8\pi^2}\,.\label{eq:flab}}
It is straightforward to solve the above 
\ee{\lambda(\mu)=\f{\lambda(\mu_0)}{1-\f{9\lambda(\mu_0)}{8\pi^2}\log({\mu}/{\mu_0})}\,;\qquad{m^2}(\mu)=\f{m^2(\mu_0)}{\big[1-\f{9\lambda(\mu_0)}{8\pi^2}\log({\mu}/{\mu_0})\big]^{1/3}}\,,\label{eq:run11}}
where the scale $\mu_0$ fixes the physical input values of the parameters which in principle are provided by the appropriate measurements.
Using the above we may easily write down the effective potential with running constants
\ea{V_{\rm eff}\left(\chi(\mu)\right)
&=\f{1}{2}m^2(\mu){\chi}^2(\mu) +\f{\lambda(\mu)}{4}{\chi}^4(\mu)\nonumber \\&+\f{\big(m^2(\mu)+3{\lambda(\mu)}{\chi}^2(\mu)\big)^2}{64\pi^2}\bigg[\log\left(\f{m^2(\mu)+3{\lambda(\mu)}{\chi}^2(\mu)}{\mu^2}\right)-\f{3}{2}\bigg]
\label{eq:Veffsca3}\,,}
where for clarity we have have denoted all $\mu$-dependence explicitly and neglected the running vacuum energy $V_\Lambda(\mu)$. 

The scale $\mu$ is usually chosen to 
match the energy scale of the process one is considering
because then the logarithms appearing in the loop corrections are generally small, and therefore the loop expansion can be expected to converge faster.

However, in the case of the effective potential, the characteristic
energy scale generally depends on the field value.
Therefore it becomes natural to make the scale $\mu$ depend on the field value $\chi$. More precisely, we define
a suitable function $\mu_*(\chi)$, chosen in such a way that the loop corrections are small.
This leads to the {\em renormalisation group improved} (RGI) effective potential,
\ea{V_{\rm RGI}(\chi)
&=\f{1}{2}m^2(\mu_*)\frac{Z(\mu_0)}{Z(\mu_*)}{\chi}^2 +\f{\lambda(\mu_*)}{4}\frac{Z(\mu_0)^2}{Z(\mu_*)^2}{\chi}^4\nonumber \\&+
\frac{1}{64\pi^2}
\left(
m^2(\mu_*)+3{\lambda(\mu_*)}\frac{Z(\mu_0)}{Z(\mu_*)}{\chi}^2\right)^2
\bigg[\log\bigg(\f{m^2(\mu_*)+3{\lambda(\mu_*)} \frac{Z(\mu_0)}{Z(\mu_*)}{\chi}^2}{\mu_*^2}\bigg)-\f{3}{2}\bigg]
\label{eq:VRGI}\,,}
where $\mu_*=\mu_*(\chi)$, by $\chi$ we denote the field defined with fixed renormalisation scale $\mu_0$, i.e., $\chi=\chi(\mu_0)$, and the field renormalisation factors are
\begin{equation}
\frac{Z^{1/2}(\mu_0)}{Z^{1/2}(\mu)}=\exp\bigg(-\int_{0}^{\log\left(\frac{\mu}{\mu_0}\right)}\gamma(t)\dd t\bigg).
\end{equation}

Here we point out that we reserve the word 'improved' for the result where $\mu$-independence is exploited in order to optimize the convergence of the perturbative expansion i.e. for (\ref{eq:VRGI}) but not for (\ref{eq:Veffsca3}) in contrast to some other works.

In simple cases, when all particle masses are proportional to the field $\chi$, it is common to choose $\mu=\chi$ \cite{Casas:1994qy}. However, in situations which involve other energy scales (for example the spacetime curvature in our case) this does not necessarily work, and one needs a 
general presciption for determining $\mu$ for each field value. 
A natural choice is to set $\mu$ in such a way that the one loop correction vanishes, as was recently advocated in \cite{Chataignier:2018aud}. In the case of Eq.~(\ref{eq:Veffsca3}), this means choosing $\mu=\mu_*$, where $\mu_*$ is given by solving the equation
\ee{\mu_*^2= 
e^{-3/2}\left[m^2(\mu_*)+3{\lambda(\mu_*)}\frac{Z(\mu_0)}{Z(\mu_*)}{\chi}^2\right]\,.\label{eq:muc}}
Even in this simple theory this equation cannot be solved
analytically, but numerically it is straightforward.
With the scale choice (\ref{eq:muc}), the RGI potential
can be written simply as
\ea{V_{\rm RGI}(\chi)
&=\f{1}{2}m^2(\mu_*)\frac{Z(\mu_0)}{Z(\mu_*)}{\chi}^2 +\f{\lambda(\mu_*)}{4}\frac{Z(\mu_0)^2}{Z(\mu_*)^2}{\chi}^4\label{eq:VRGI0}\,.}

We can now repeat the process of RG improvement for the case of an arbitrary curved spacetime. Note that at no point in our derivation have we used an expansion assuming the background curvature to be a small correction on top of the flat space result. This is important as the case of significant background curvature is precisely the relevant one for many applications. The only assumption is that the quantized matter field is energetically sub-dominant, which means that it can be treated as a spectator field on a curved classical background \cite{Markkanen:2017dlc}.

In addition to (\ref{eq:flab}) in curved spacetime $\beta$-functions related to the gravitational terms (\ref{eq:treecurve}) emerge, which are easy to derive by using (\ref{eq:CS}) for the curved spacetime effective potential (\ref{eq:curve2}), which results in
\ea{\beta_\xi&=\f{6\lambda(\xi-1/6)}{16\pi^2}\,;&
\beta_{V_\Lambda}&=\f{m^4/2}{16\pi^2}\,; &\beta_\kappa&=-\f{m^2(\xi-1/6)}{16\pi^2}\,;& \nonumber \\\beta_{\alpha_1}&=\f{(\xi-1/6)^2/2}{16\pi^2}\,; &\beta_{\alpha_2}&=-\f{1/180}{16\pi^2}\,;& \beta_{\alpha_3}&=\f{1/180}{16\pi^2}\,. \label{eq:betaC}}
The above could easily be solved and used to improve the perturbative result (\ref{eq:curve2}) very similarly to the flat space case in (\ref{eq:Veffsca3}). 

From a calculational point of view RG improvement in curved spacetime proceeds as a very natural extension of the usual steps made in flat space. There are however very important qualitative differences that arise in the presence of background curvature. For example, in the simple scalar field case in flat space we optimized the expansion in the improved potential (\ref{eq:Veffsca3}) by making the logarithms small with the choice (\ref{eq:muc}). Due to the $R$-dependence in the logarithm in (\ref{eq:curve2}) the analogous choice in curved spacetime would in be
\ee{\mu_*^2=
e^{-3/2}\left|m^2(\mu_*)+3{\lambda(\mu_*)}\frac{Z(\mu_0)}{Z(\mu_*)}{\chi}^2+
\left(\xi(\mu_*)- \frac{1}{6}\right)R
\right|\,.}
This is a generic feature that arises in curved spacetimes: whenever we optimize the expansion in the improved potential we unavoidably introduce a curvature dependence in the running scale i.e. it leads to \textit{curvature induced running} of the parameters.
In the case of high background curvature, $R\gg m^2(\mu)+3{\lambda(\mu)}{\chi}^2$, the dominant contribution to the running of parameters comes from the scalar curvature, an effect which obviously is completely missed when using a flat space approximation for the potential. In the context of a potential electroweak vacuum instability in the early Universe this effect was first discovered in \cite{vacstab7}. 

Another important feature with qualitatively significant consequences is the generation of gravitational operators that couple to the scalar field, which we have already mentioned on several occasion. In addition to the well-known non-minimal coupling $\xi$ leading to a tree-level coupling between curvature and the scalar field, the $\mathcal{O}(R^2)$-type operators visible in (\ref{eq:curve2}) introduce potentially significant modifications in the effective potential. As seen from (\ref{eq:betaC}) all these operators are generically radiatively induced. 

Putting all this together, we can write the final expression for the RGI effective potential, including the gravitational terms, as:
\begin{align}
V_{\rm{RGI}}(\chi) =& \frac{1}{2}\left(m^2(\mu_{*}(\chi))+\xi(\mu_{*}(\chi))R\right)\frac{Z(\mu_0)}{Z(\mu_{*}(\chi))}\chi^2 + \frac{Z^2(\mu_0)}{Z^2(\mu_{*}(\chi))}\frac{\lambda(\mu_{*}(\chi))}{4}\chi^4 \nonumber\\ 
&+ V_{\Lambda}(\mu_{*}(\chi)) - \kappa(\mu_{*}(\chi)) R + \alpha_1(\mu_{*}(\chi)) R^2 + \alpha_2(\mu_{*}(\chi)) R_{\mu\nu}R^{\mu\nu}\nonumber\\
&+ \alpha_3(\mu_{*}(\chi))R_{\mu\nu\delta\eta}R^{\mu\nu\delta\eta}
+ \frac{\mathcal{M}^4(\chi)}{64\pi^2}\left[\log\left(\frac{|\mathcal{M}(\chi)^2|}{\mu_{*}^2(\chi)}\right) - \frac{3}{2}\right]\nonumber\\
&+ \frac{\frac{1}{90}\left(R_{\mu\nu\delta\eta}R^{\mu\nu\delta\eta} - R_{\mu\nu}R^{\mu\nu}\right)}{64\pi^2}\left[\log\left(\frac{|\mathcal{M}^2(\chi)|}{\mu_{*}^2(\chi)}\right)\right],
\end{align}
where:
\begin{equation}
\mathcal{M}^2(\chi) = m^2(\mu_{*}(\chi)) + 3\lambda(\mu_{*}(\chi))\frac{Z(\mu_0)}{Z(\mu_{*}(\chi))}\chi^2 + \left(\xi(\mu_{*}(\chi)) - \frac{1}{6}\right)R,
\end{equation}
$\chi$ is the field renormalized at $\mu_0$, and $\mu_{*}(\chi)$ is the chosen scale for the renormalization group improvement, as a function of $\chi$.

Now we can proceed to discuss the derivation of the curved spacetime effective potential for more involved theories containing fermions and gauge fields in addition to scalars. Despite a substantial increase in calculational work qualitatively the main modifications one encounters are the same as in the simple scalar field case just discussed. 
\section{Effective potential via the Heat Kernel}
\label{eq:appa}
In this section we present the steps for the derivation of an effective potential containing  scalar, fermion and gauge fields with the proper gauge-fixing terms at the ultraviolet limit. Although the main application we have in mind is the specific case of the SM on a homogeneous and isotropic spacetime relevant for cosmological applications, the results are presented in a general form applicable for arbitrary field content and an arbitrary background metric. We also note that many of the slightly formal results of this section become clearer when they are implemented in practice, which we will do in section \ref{sec:SM} for the SM in de Sitter space. 

Following the standard discussion found in  \cite{Peskin:1995ev} first we briefly review the use of functional determinants in the calculation the 1-loop effective potential in flat space.
The generating functional defined via a path integral in flat space reads
\ee{Z[J]=\int \mathcal{D}\varphi~e^{iS[\varphi]+i\int d^4x~J\varphi},\label{eq:z}}
where $\varphi$ is some generic scalar field with possibly non-trivial group structure and $J$ is the usual source term.

By performing a Legendre transformation as \ee{{\Gamma[\varphi]=-i\log Z[J]-\int d^4x~J(x)\varphi(x)}\, ,\label{eq:gamma0}}
one obtains the effective action $\Gamma[\varphi]$  and, for a constant field, $\varphi$ the effective potential as 
\ee{\Gamma[\varphi]\equiv\Gamma^{(0)}[\varphi]+\Gamma^{(1)}[\varphi]+\cdots\equiv \int d^4x\, {\cal L}_{\rm eff}~\overset{\varphi\,=\,{const.}}{=}-V_{\rm eff}(\varphi)\int d^4x\,.}

The standard approach in the 1-loop approximation, which contains only terms quadratic in fluctuations, comes by using the path integral generalization of the formulae
\ea{\int e^{-x^i A_{ij}x^j }dx&\propto \f{1}{\sqrt{\det A_{ij}}}\,,\int e^{-x^i A_{ij}x^*{}^j }dxdx^*\propto \f{1}{\det A_{ij}}\,,\nonumber \\\int e^{-\bar{c}^i A_{ij}c^j }d\bar{c}dc&\propto {\det A_{ij}}\,;\quad \{c_i,c_j\}=0\,,}
to write the 1-loop contribution to the effective potential as a sum of 'tracelogs'
\ee{\Gamma^{(1)}[\varphi]=i\sum_kn_k\log\det D_k =i\sum_kn_k\,{\rm Tr}\log D_k\, .\label{eq:gammaex2}}
with the pre-factors determined by the field content and the group structure of the theory in question.

The above summarizes the standard approach for the derivation of effective action to 1-loop order, which can also be generalized for the case of a curved spacetime. Symbolically nothing changes for the tracelogs, but in curved spacetime finding explicit expressions for them becomes non-trivial due replacement of $\partial_\mu$ with the $\nabla_\mu$-operator containing the spacetime connection 
and in the integral over spacetime indices the inclusion of the measure $\sqrt{-g}$.

Following the steps and the conventions of \cite{ParkerToms} we can express any effective action  
given in terms of a tracelog as an integral over a fictitious time parameter $\tau$ often called proper time,
\ee{
\Gamma^{(1)}[\varphi]=\f{i}{2}{\rm Tr}\log\,D= -\f{i}{2}{\rm Tr}\left[ \int_0^\infty \f{d\tau}{\tau} e^{-i\tau D} \right]
= -\f{i}{2} \int d^4 x \sqrt{-g} \int_0^\infty \f{d\tau}{\tau}
 {\rm tr}\left[K(\tau;x,x)\right]\,,
\label{eff_pot_kernel}
} 
where $D$ is yet an unknown matrix,  $"{\rm Tr}"$ denotes a trace over all indices, $"{\rm tr}"$ a trace with the spacetime coordinates excluded and
\ee{
K(\tau; x,y) \equiv e^{-i\tau D}\,,\label{eq:kern}
}
is the \textit{heat kernel}. Note that 
no assumptions of the underlying theory have been made when writing (\ref{eff_pot_kernel}) and (\ref{eq:kern}), except perhaps that the integral is well-defined. 

The efficiency of heat kernel formalism lies in the existence of approximation schemes allowing one to perform the proper time integral in (\ref{eff_pot_kernel}) and hence obtain a solution for the effective action:
if the matrix $D$ can be expressed as
\ee{
 D = \Box + X\,,\label{eq:D}
}
where $X$ is an arbitrary matrix in spin and group degrees of freedom, a small $\tau$ expansion of the heat kernel  becomes analytically tractable. It will turn out that to 1-loop order scalars, fermions and gauge fields all can be manipulated to have the form (\ref{eq:D}) even in the presence of background curvature. Note that the $\Box$-operator contains the metric connection (and possibly the spin connection) via the $\nabla_\mu$-operator, as is described in section 3.8 of \cite{Birrell:1982ix} for fields with any spin.

The small $\tau$ expansion of the heat kernel corresponds to the UV  or local limit and has been studied and implemented in the physical context by a number of people over the years. Notable early work on heat kernel techniques may be found in \cite{Schwinger:1951nm,DeWitt:1965jb,Seeley:1967ea,Gilkey:1975iq,Minakshisundaram:1949xg,Hadamard} and see \cite{Avramidi:2000bm} for a review. 

In \cite{Parker:1984dj,Jack:1985mw} a form of the heat kernel that sums all scalar curvature contributions in the small proper time approximation was provided. This is obtained via the ansatz
\ee{
K(\tau; x,x) = \f{i}{(4\pi i\tau)^{n/2}}
\exp\left[-i\tau\Big( X - \f{1}{6}R\Big)\right]\Omega(\tau)\,,
\label{kernel_exp}
}
where we have analytically continued the dimensions to $n$ and where
\ee{
\Omega(\tau) = \sum\limits_{k=0}^\infty a_k (i\tau)^k\,;\qquad a_0 = 1\,,\qquad a_1 = 0\,,
}
with
\ee{
a_2= -\f{1}{180}R_{\mu\nu}R^{\mu\nu}+\f{1}{180}R_{\mu\nu\rho\sigma}R^{\mu\nu\rho\sigma}-\f{1}{30}\Box R+\f{1}{6}\Box X+\f{1}{12}W_{\alpha\beta}W^{\alpha\beta}
\,.\label{eq:SDa2}}
The $W_{\alpha\beta}$ is defined as 
\ee{W_{\alpha\beta}\Psi=\big[\nabla_\alpha,\nabla_\beta\big]\Psi\,,} and depends on spacetime structure of the particular field contribution $\Psi$ to the effective potential i.e. it will be different for scalars, fermions and gauge fields as will be illustrated in sections \ref{sec:sca} -- \ref{sec:scg}. It is precisely the form (\ref{kernel_exp}) that turns out to correspond to the elementary UV derivation of section \ref{sec:effp}. 

We can now write down the effective action (\ref{eff_pot_kernel}) by using eqs.~(\ref{kernel_exp}) -- (\ref{eq:SDa2}) and performing the integral over proper time in $n$ spacetime dimensions,
\ea{
\Gamma^{(1)}[\varphi] &= \f{1}{2(4\pi)^{n/2}}  \int d^n x \sqrt{-g}\;{\rm tr}\left[
 \left(\f{\mathcal{M}^2}{\mu^2}\right)^{\f{n}{2}-2} \sum\limits_{k=0}^\infty \left(\mathcal{M}^2\right)^{2-k} a_k
 \Gamma\left(k-\f{n}{2}\right)\right]
\nonumber\\[2mm]
&= -\int d^n x \sqrt{-g}\;\f{1}{64\pi^2}\,{\rm tr}\left[\mathcal{M}^4\left(\log\f{|\mathcal{M}^2|}{\tilde\mu^2}
 - \f{3}{2} \right)+2a_2\log\f{|\mathcal{M}^2|}{\tilde\mu^2}\right] + \ldots 
\label{eff_pot_kernel_final}
}
where we have only included the leading logarithmic terms, denoted the curved spacetime effective mass as 
\ee{
\mathcal{M}^2 \equiv  X - \f{1}{6}R\label{eq:cm}\,,
}
and absorbed the divergences in the renormalization scale as
\ee{
\log\tilde\mu^2 = \log\mu^2 + \f{2}{4-n} - \gamma_E + \log(4\pi)\,.
\label{mu-tilde}
}
Note that all group and spacetime indices are present implicitly. In flat space in the $\overline{\rm MS}$ renormalization scheme the last three terms on the RHS of  (\ref{mu-tilde}) are subtracted by the counter terms after the full expression is expanded in the limit $n \to 4$ resulting in the replacement $\log\tilde\mu^2 \rightarrow \log\mu^2$. We will also make use of this replacement although in curved space non-logarithmic finite pieces containing curvature dependence are generated as the result of the interplay between the pole $\propto (4-n)^{-1}$ and $n$-dependent terms in $R,R_{\mu\nu}$ and $R_{\mu\nu\rho\sigma}$. These terms we absorb in $\xi$, $\kappa$, $V_{\Lambda}$ and the $\alpha$'s in the tree level action (\ref{eq:actS}) and (\ref{eq:treecurve})\footnote{For the conformal anomaly this is of course not possible, however this contribution does not couple to the Higgs and hence in the subsequent discussion can be ignored.}. 

In what follows we use the formula (\ref{eff_pot_kernel_final}) to compute the one-loop contributions to the effective potential from scalar, fermion and gauge fields. For the scalar case we can easily see that the first term in the expansion (\ref{eff_pot_kernel_final}) corresponds with the elementary derivation (\ref{eq:curve}) in section (\ref{eq:1cS}).  Like in (\ref{eq:curve}) we have only included the real part of the result as the infrared modes potentially giving a complex result are not included in the local (UV) expansion (\ref{kernel_exp}) with (\ref{eq:SDa2}). Finally, we choose to drop the $\Box$-type terms in (\ref{eq:SDa2}) as they will not give rise to divergences or $\mu$-dependence when $\mu$ is a constant as in $\overline{\rm MS}$, which follows from the assumption of an unbounded Universe and partial integration.

\subsection{Scalar}
\label{sec:sca} 
Since scalar fields will always result in a tracelog given via an operator of the form (\ref{eq:D}), one may directly implement the expression in (\ref{eff_pot_kernel_final}). Much like in (\ref{eq:trlog1}) parametrizing the 1-loop contribution to the effective action via an effective mass parameter now denoted as $m_s$ with a non-minimal coupling gives
\ee{\Gamma^{(1)}_s[\varphi] = \f{i}{2}{\rm Tr}\log\left[\Box + m_s^2+\xi R\right]\,.\label{eq:trlog}}
The effective mass in (\ref{eq:cm}) then becomes simply
\ee{\mathcal{M}^2_s= m_s^2+\bigg(\xi-\f{1}{6}\bigg)R\label{eq:Ms}\,,}
where "s" stands for scalar and similarly the relevant higher order curvature terms are
\ee{a_{2,s}=-\f{1}{180}R_{\mu\nu}R^{\mu\nu}+\f{1}{180}R_{\mu\nu\rho\sigma}R^{\mu\nu\rho\sigma}\,,\label{eq:as}}
where we used that fact that for a scalar $\phi$ one has
\ee{\big[\nabla_\mu,\nabla_\nu]\phi=0\quad \Rightarrow \quad W_{\mu\nu}=0\,.}
Here we further emphasize that all scalars have the same $a_2$-contribution.
\subsection{Fermion}
\label{sec:scf}
From the SM Lagrangian (\ref{eq:quad_action}) one sees that a typical fermion contribution needs some work before (\ref{eff_pot_kernel_final}) can be used, since it is not in the form (\ref{eq:D}), but
\ee{\Gamma^{(1)}_f[\varphi]=-i\,{\rm Tr}\log\big[i\nabla_\mu \gamma^\mu-m_f\big]\,,}
where much like for scalars, the subscript "$f$" stands for fermion.
Using very similar steps as in flat space i.e. the fact that $\log\det = {\rm Tr}\log$ and that 
\ea{\det\left[i\nabla_\mu \gamma^\mu-m_f\right]&=\left[\det(i\nabla_\mu \gamma^\mu-m_f)\det(\gamma_5\gamma_5)(i\nabla_\mu \gamma^\mu-m_f)\right]^{1/2}\nonumber \\&=[\det(-i\nabla_\mu \gamma^\mu-m_f)(i\nabla_\mu \gamma^\mu-m_f)]^{1/2}\,,}
with the help of the relation $(\gamma^\mu\nabla_\mu)^2=\Box+R/4$ leads to
\ee{\Gamma^{(1)}_f[\varphi]=-\f{i}{2}{\rm Tr}\log\big[\Box + m_f^2+R/4\big]\,,}
and the fermionic effective mass (\ref{eq:cm}) in curved spacetime 
\ee{\mathcal{M}^2_f= m_f^2 +\f{R}{12}\label{eq:Mf}\,.}
Unlike for the scalar case the $W_{\mu\nu}$ term from (\ref{eq:SDa2}) gives a no-zero contribution due to the spin connection of the fermion. By using \cite{Jack:1985mw}
\ee{\big[\nabla_\mu,\nabla_\nu]\psi=-\f{1}{4}R_{\mu\nu\alpha\beta}\gamma^\alpha\gamma^\beta\psi=W_{\mu\nu}\psi\,,}
for Dirac fermions $\psi$ in curved spacetime and familiar identities from trace technology for the combination of four $\gamma$-matrices we can write
\ea{{\rm tr }\{ a_{2,f}\}&={\rm tr }\{\mathbb{1}_{\rm Group} \}\bigg [-\f{1}{45}R_{\mu\nu}R^{\mu\nu}-\f{7}{360}R_{\mu\nu\rho\sigma}R^{\mu\nu\rho\sigma}\bigg ]\,.\label{eq:a2f}}
In the above we have performed the trace over Dirac indices, but for completeness left in the trivial trace over any group indices, which for example for the SM quarks simply gives an overall factor of 3 from the three different colors and a factor of 1 for leptons.
\subsection{Gauge}
\label{sec:scg}
The gauge contributions to the effective potential are the most non-trivial to calculate due to the explicit dependence on $R_{\mu\nu}$ and $\nabla^\mu\nabla^\nu$ as is visible from (\ref{eq:quad_action}) for the SM\footnote{For a similar derivation we refer the reader to section 7.9 of \cite{ParkerToms}.}.

The main difficulty comes from gauge fixing. We use the the so-called $R_\xi$ or background ('t Hooft) gauges \cite{Peskin:1995ev} to fix the gauge, which in order not to create confusion with the non-minimal coupling we parametrize with $\zeta$. Generically, gauge fields give rise to contributions of the form
\ee{\Gamma^{(1)}_g[\varphi] = \f{i}{2}{\rm Tr}\log\left[\Box g^{\mu\nu}+\left(\f{1}{\zeta}-1\right)\nabla^\mu\nabla^\nu+m^2_gg^{\mu\nu}+R^{\mu\nu}\right]\,.\label{eq:trlogZ}}
The above can be simplified by first splitting the vector into scalar and orthogonal components as
\ee{A^\mu =A^\mu_\bot+\nabla^\mu A\,; \qquad \nabla_\mu A^\mu_\bot=0\,,}
with which and the help of standard commutator formula \ee{\big[\nabla_\mu,\nabla_\nu\big]A^\rho=-R^\rho_{~\,\alpha\mu\nu}A^\alpha\,,\label{eq:comma}}
 we can write
\ea{&\bigg[\Box g_{\nu\mu}+\left(\f{1}{\zeta}-1\right)\nabla_\nu\nabla_\mu+m^2_gg_{\nu\mu}+R_{\nu\mu}\bigg]A^\mu 
\nonumber \\ = &\big[\Box g_{\nu\mu}+m^2_gg_{\nu\mu}+R_{\nu\mu}\big]A^\mu_\bot+\zeta^{-1}\big[\Box g_{\nu\mu}+\zeta m^2_gg_{\nu\mu}+R_{\nu\mu}\big]\nabla^\mu A\,.\label{eq:gexp}}
From this it follows that the gauge tracelog splits into two separate pieces. The first one is almost of the form required by the hear kernel results, were it not the orthogonality constraint "$\bot$". Its effect can be deduced by studying the unconstrained eigenvalue equation
\ea{\big[\Box g_{\mu\nu}+m^2_gg_{\mu\nu}+R_{\mu\nu}\big]A^\nu&=\lambda A_\mu\qquad\qquad  \big\vert\cdot \nabla^\mu\nonumber \\ \Leftrightarrow\quad \left[\Box +m^2_g\right]\nabla_\nu A^\nu&=\lambda\nabla_\nu A^\nu\,.}
These eigenvalues have to be removed from the constrained ones resulting from $A^\mu_\bot$, giving us symbolically the relation
\ea{&{\rm eigenvalues}\Big\{\big[\Box g_{\mu\nu}+m^2_gg_{\mu\nu}+R_{\mu\nu}\big]A^\nu_\bot\Big\}\nonumber \\ =~&{\rm eigenvalues}\Big\{\big[\Box g_{\mu\nu}+m^2_gg_{\mu\nu}+R_{\mu\nu}\big]A^\nu_{\phantom{{\bot}}}\Big\}-{\rm eigenvalues}\Big\{\left[\Box +m^2_g\right]\nabla_\nu A^\nu\Big\}\,,}
where the last piece is simply a scalar contribution.

The second term in (\ref{eq:gexp}) we can evaluate by invoking consistency: at the limit $\zeta=1$ all terms resulting from the $\nabla^\mu\nabla^\nu$ piece in (\ref{eq:trlogZ}) must vanish, which completely fixes the remaining contribution allowing us to write for the gauge field
\ee{\Gamma^{(1)}_g[\varphi] = \f{i}{2}{\rm Tr}\log\left[\Box g^{\mu\nu}+m^2_gg^{\mu\nu}+R^{\mu\nu}\right]-\f{i}{2}{\rm Tr}\log\left[\Box+m^2_g\right]+\f{i}{2}{\rm Tr}\log\left[\Box+\zeta m^2_g\right]\,,\label{eq:gauG}}
where only the first operator is a matrix in terms of spacetime indices and the other two are scalars.

As one may see from (\ref{eq:gauG}) a gauge field will result in 3 separate pieces with the effective masses from (\ref{eq:cm}) 
\ee{\big(\mathcal{M}^2_g\big)^{\mu\nu}= m_g^2g^{\mu\nu} + R^{\mu\nu}-\f{g^{\mu\nu}}{6}R\,;\qquad\mathcal{M}^2_{g,s}= m_g^2-\f{R}{6}\,;\qquad\mathcal{M}^2_{g,\zeta s}= \zeta m_g^2-\f{R}{6}\label{eq:Mg}\,.}
For gauge fields we can again use (\ref{eq:comma})
\ee{\big[\nabla_\mu,\nabla_\nu\big]A^{\alpha}=-R_{\mu\nu~\beta}^{~~\,\alpha}A^\beta=(W_{\mu\nu})^\alpha_{~\,\beta}A^\beta\quad\Rightarrow\quad(W_{\mu\nu})_{\alpha}^{~\,\delta}(W^{\mu\nu})_{\delta}^{~\,\beta}=R_{\mu\nu\alpha}^{~~~~\delta}R^{\mu\nu~~\beta}_{~~~\delta}\,,}
for obtaining an expression for the $a_2$-contribution for the first term on the right hand side of (\ref{eq:gauG})
\ee{(a_{2,g})_\alpha^{~\,\beta}=-\f{g_\alpha^{~\,\beta}}{180}R_{\mu\nu}R^{\mu\nu}+\f{g_\alpha^{~\,\beta}}{180}R_{\mu\nu\rho\sigma}R^{\mu\nu\rho\sigma}+\f{1}{12}R_{\mu\nu\alpha}^{~~~~\delta}R^{\mu\nu~~\beta}_{~~~\delta}\,,\label{eg:a2g}}
with "$g$" for gauge and the remaining two scalar pieces trivially give rise to two contributions as in (\ref{eq:as}).
\section{The Standard Model}
\label{sec:SM}
By making use of the results we derived in the previous section we present the steps for calculating the effective potential for the SM Higgs to 1-loop order in curved spacetime. We will show the explicit derivation  in a set-up including only the Higgs doublet, the massive vector bosons $W^\pm$ and $Z^0$ and the top quark, as from this result a generalization which includes the complete particle content of the SM is straightforward. 

We start with the Lagrangian
\ee{
{\cal L}_{\rm SM} ={\cal L}_{\rm YM} + {\cal L}_{\rm F} + {\cal L}_{\Phi}+{\cal L}_{GF}+{\cal L}_{GH}+\cdots
\label{eq:lag}}
with
\ea{
{\cal L}_{\rm YM} &= -\f{1}{4}\left(F^a_{\mu\nu}\right)^2-\f{1}{4}\left(F_{\mu\nu}\right)^2
+\cdots\,; \qquad\qquad\qquad\qquad\qquad a =1,2,3\\\label{eq:higgsl}
{\cal L}_{\Phi} &= \left(D_\mu \Phi \right)^\dagger \left(D^\mu \Phi \right) +m^2\Phi^\dagger \Phi-\xi R\Phi^\dagger \Phi- \lambda (\Phi^\dagger \Phi)^2
\,;\\
{\cal L}_{\rm F} &= \bar Q_L i\gamma^\mu D_\mu Q_L + \bar t_R i\gamma^\mu D_\mu t_R
 + \left(-y_t \bar Q_L(i\sigma^2)\Phi^* t_R + {\rm h.c.}\right)+\cdots\,;
}
where $\Phi$ and $Q_L$ are the Higgs and the left-handed top/bottom doublets, respectively
\ee{\Phi = \f{1}{\sqrt{2}}\left(\begin{array}{c} -i(\chi_1 - i \chi_2) \\ \varphi+ (h + i \chi_3)\end{array}\right)\,;\qquad\qquad\quad Q_L=\left(\begin{array}{c} t\\ b\end{array}\right)_L\,, }
and where $\varphi$ is the vacuum expectation value and finally the $\chi_i$'s are the would be Goldstone bosons. Since our calculation is performed in curved space the covariant derivative, in addition to the gauge connection, contains a metric dependence via the covariant $\nabla_\mu$
\ea{
D_\mu &= \nabla_\mu - ig \tau^a A^a_\mu - ig'Y A_\mu;\qquad \tau^a=\sigma^a/2\, .
}
Also note that the $\gamma$-matrices satisfy the curved space generalization of the usual relation
\ee{\big\{\gamma_\mu,\gamma_\nu\big\}=g_{\mu\nu}\,,} which as shown in section \ref{sec:scf} plays a role in the contributions from the fermions. The last two terms in (\ref{eq:lag}), ${\cal L}_{GF}$ and ${\cal L}_{GH}$, are the gauge fixing and ghost contributions. As discussed in section (\ref{sec:scg}) we will use the background gauge where the gauge parameters are left unspecified. With these choices the gauge fixing Lagrangian can then be written as
\ee{{\cal L}_{\rm GF}=-\f{1}{2}G^2\,;\qquad G^j=\f{1}{\sqrt{\zeta^j}}\left(\nabla^\mu A^j_\mu-\zeta^j E^{ja}\chi^a\right);\qquad j=1,2,3,4;\qquad a=1,2,3\, ,\label{eq:gf}}
with the definitions 
\ee{E^{ia}\equiv\begin{pmatrix}
g&\phantom{-}0&\phantom{-}0\\0&\phantom{-}g&\phantom{-}0\\0&\phantom{-}0&\phantom{-}g\\0&\phantom{-}0&-g'
\end{pmatrix}\,;\qquad A^4_\mu\equiv A_\mu.\label{eq:crosst}}
Finally, since we are fixing gauge in a non-Abelian gauge theory we must also introduce a ghost term. The ghost Lagrangian can be written from
\ee{\mathcal{L}_{\rm GH}=\bar{c}^i\left(\f{\delta G^i}{\delta\alpha^j}\right)c^j\,,}
where $G^i$ is the gauge fixing function from (\ref{eq:gf}), the $c$'s are the ghost fields and the $\alpha^i$ the parameters of the gauge transformations. 

Now we may write the Lagrangian to quadratic order in fluctuations; it requires some algebra but is a straightforward exercise. At this stage the only complication arising from having a curved background is the fact that covariant derivatives for the gauge fields do not commute as given in (\ref{eq:comma}). Taking this into account one gets\footnote{with the usual mass eigenstates
\ee{W^\pm_\mu=\f{1}{\sqrt{2}}\left(A^1_\mu\mp i A^2_\mu\right)\,;\qquad Z^0_\mu=\f{1}{\sqrt{g^2+(g')^2}}\left(g A^3_\mu -g'A_\mu\right)\,,\nonumber}}
\ea{
{\cal L}_{\rm SM} &=\f{m^2}{2}\varphi^2-\f{\lambda}{4}\varphi^4-\f{1}{2}h\left[\Box+m^2_h+\xi R\right]h+\bar t\left[i\nabla_\mu \gamma^\mu-m_t\right]t\nonumber \\&+W^+_\mu\left[\Box g^{\mu\nu}+\left(\f{1}{\zeta_W}-1\right)\nabla^\mu\nabla^\nu+m_W^2g^{\mu\nu}+R^{\mu\nu}\right]W^-_\nu\nonumber \\&+\f{1}{2}Z^0_\mu\left[\Box g^{\mu\nu}+\left(\f{1}{\zeta_Z}-1\right)\nabla^\mu\nabla^\nu+m^2_Zg^{\mu\nu}+R^{\mu\nu}\right]Z^0_\nu\nonumber \\
&-\f{1}{2}\sum_{a=1,2} \chi^a\left[\Box +\zeta_W m_W^2+m^2_\chi+\xi R\right]\chi^a-\f{1}{2}\chi^3\left[\Box +\zeta_Z m_Z^2+m^2_\chi+\xi R\right]\chi^3\nonumber \\&-\sum_{a=1,2} \bar{c}^a\left[\Box +\zeta_W m_W^2\right]c^a-\bar{c}^3\left[\Box +\zeta_Z m_Z^2\right]c^3+\cdots\,,
\label{eq:quad_action}
}
where we have chosen separate gauge fixing parameters for the $W^\pm$ and $Z^0$ contributions and defined the mass parameters
\ea{
m^2_h&=-m^2+3\lambda\varphi^2,\quad \,m^2_t=\f{y_t^2}{2}\varphi^2,\quad m^2_W = \f{g^2}{4}\varphi^2\,,\nonumber \\
m^2_Z &= \f{g^2 + (g')^2}{4}\varphi^2\,,\quad m^2_\chi=-m^2+\lambda\varphi^2
\,.\label{eq:masses}}

For an intermediate result we use the steps shown in (\ref{sec:sca}), (\ref{sec:scf}) and (\ref{sec:scg}) of section \ref{eq:appa} to write (\ref{eq:quad_action}) in terms of tracelogs that are calculable with the heat kernel technology. Explicitly this gives
\ea{V^{(1)}_{\rm SM}(\varphi)=&-\f{i}{2}{\rm Tr}\log\big[\Box+m^2_h+\xi R\big]+\f{i}{2}{\rm Tr}\log\big[\Box+m_t^2+{R}/{4}\big]\nonumber \\&-i\,{\rm Tr}\log\left[\Box g_{\mu\nu}+m_W^2 g_{\mu\nu}+R_{\mu\nu}\right]+i\,{\rm Tr}\log\left[\Box+m^2_W\right]\nonumber \\&-\f{i}{2}{\rm Tr}\log\left[\Box g_{\mu\nu}+m_Z^2 g_{\mu\nu}+R_{\mu\nu}\right]+\f{i}{2}{\rm Tr}\log\left[\Box+m^2_Z\right]\nonumber \\&-i\,{\rm Tr}\log\left[\Box +\zeta_W m_W^2+m^2_\chi+\xi R\right]-\f{i}{2}{\rm Tr}\log\left[\Box +\zeta_Z m_Z^2+m^2_\chi+\xi R\right]\nonumber \\&+i\,{\rm Tr}\log\left[\Box +\zeta_W m_W^2\right]+\f{i}{2}{\rm Tr}\log\left[\Box +\zeta_Z m_Z^2\right]+\cdots\,,\label{eq:tralgsm}}
where the definitions for the mass parameters can be found in (\ref{eq:masses}). Note that the $\zeta$-dependent mass terms given by (\ref{eq:Mg}) for the $W^\pm_\mu$ and $Z^0_\mu$ fields have been canceled by the ghost contribution.

It proves convenient to split (\ref{eq:tralgsm}) into scalar, fermion and gauge contributions as
\ee{V^{(1)}_{\rm SM}(\varphi)=V^{(1)}_{{\rm SM}}(\varphi)_{scalar}+V^{(1)}_{{\rm SM}}(\varphi)_{fermion}+V^{(1)}_{{\rm SM}}(\varphi)_{gauge}\,.}
Collecting all the scalar pieces and using (\ref{eff_pot_kernel_final}) along with section \ref{sec:sca} one gets

\ee{V^{(1)}_{{\rm SM}}(\varphi)_{scalar}= \sum_{\sigma=scalars}\f{n_\sigma}{64\pi^2}\left[\mathcal{M}^4_\sigma\left(\log\f{|\mathcal{M}^2_\sigma|}{\mu^2}
 - \f{3}{2} \right)+2a_{2,s}\log\f{|\mathcal{M}^2_\sigma|}{\mu^2}\right]
\label{eff_pot_scalar}
\,,}
which specifically for the Lagrangian in (\ref{eq:lag}) contains the expressions
\ea{h\,;\qquad\,\,\,\,\,\,\, n_h&=+1\,, &\mathcal{M}^2_{h}\,\,=~&m^2_h\,\,+\bigg(\xi-\f{1}{6}\bigg)R\,,\nonumber \\
W^\pm\,;\qquad\,\, n_{W,s}&=-2\,, &\mathcal{M}^2_{W,s}=~&m^2_{W}-\f{R}{6}\,,\nonumber \\
Z^0\,;\qquad\,\,\, n_{Z,s}&=-1
\,, &\mathcal{M}^2_{Z,s}=~&m^2_{Z}-\f{R}{6}\,,\nonumber \\
\chi^1\,,\chi^2\,;\qquad n_{\chi,W}&=+2\,,&\mathcal{M}^2_{\chi,W}=~&\zeta_W m^2_{W}+m^2_\chi-\f{R}{6}\,,\nonumber \\
\chi^3\,\,;\qquad \,\,n_{\chi,Z}&=+1\,, &\mathcal{M}^2_{\chi,Z}=~&\zeta_Z m^2_{Z}+m^2_\chi-\f{R}{6}\,,\nonumber \\
c^1\,,c^2\,;\qquad\, n_{c,W}&=-2\,,&\mathcal{M}^2_{c,W}=~&\zeta_W m^2_{W}-\f{R}{6}\,,\nonumber \\
c^3\,;\qquad \,\,n_{c,Z}&=-1\,, &\mathcal{M}^2_{c,Z}=~&\zeta_Z m^2_{Z}-\f{R}{6}\,,}
with the $a_{2,s}$ given by (\ref{eq:as}).

The contribution from the top quark is straightforward to express via (\ref{eff_pot_kernel_final}) with the help of section \ref{sec:scf}
\ee{V^{(1)}_{{\rm SM}}(\varphi)_{fermion,t}= -\f{1}{64\pi^2}\left[4\,{\rm tr }\{\mathbb{1}_{\rm Group} \}\mathcal{M}^4_t\left(\log\f{|\mathcal{M}^2_t|}{\mu^2}
 - \f{3}{2} \right)+2\,{\rm tr }\{ a_{2,f}\}\log\f{|\mathcal{M}^2_t|}{\mu^2}\right]
\label{eff_pot_fermion}
\,,}
where we have explicitly calculated the trace over Dirac indices, the effective mass can be found in (\ref{eq:Mf}), ${\rm tr }\{ a_{2,f}\}$ is given by (\ref{eq:a2f}) and due to color in the SM ${\rm tr }\{\mathbb{1}_{\rm Group} \}=3$ for the top  quark (${\rm tr }\{\mathbb{1}_{\rm Group} \}=1$ for leptons).

Finally we can address the contributions coming from gauge fields that have non-trivial structure in terms of spacetime/Lorentz indices. Unsurprisingly, this is the most complicated piece, which we can evaluate with the help of section \ref{sec:scg}:
\ea{&V^{(1)}_{{\rm SM}}(\varphi)_{gauge}\equiv  V^{(1)}_{{\rm SM}}(\varphi)_{gauge,W}+V^{(1)}_{{\rm SM}}(\varphi)_{gauge,Z}\nonumber +\cdots\\&= \f{2}{64\pi^2}\bigg[(\mathcal{M}^2_W)_\alpha^{\,~\beta}(\mathcal{M}^2_W)_\beta^{\,~\nu}\left(\log\f{|(\mathcal{M}^2_W)_\nu^{~\alpha}|}{\tilde\mu^2}
 - g_\nu^{~\alpha}\f{3}{2} \right)+2(a_{2,g})_\alpha^{\,~\beta}\log\f{|(\mathcal{M}^2_W)_\beta^{~\alpha}|}{\tilde\mu^2}
 \bigg]\nonumber\\&\,+ \f{1}{64\pi^2}\bigg[(\mathcal{M}^2_Z\,)_\alpha^{\,~\beta}(\mathcal{M}^2_Z\,)_\beta^{\,~\nu}\,\left(\log\f{|(\mathcal{M}^2_Z\,)_\nu^{~\alpha}|}{\tilde\mu^2}
 - g_\nu^{~\alpha}\f{3}{2} \right)\,+2(a_{2,g})_\alpha^{\,~\beta}\log\f{|(\mathcal{M}^2_Z\,)_\beta^{~\alpha}|}{\tilde\mu^2}
 \bigg]+\cdots\,,
\label{eff_pot_gauge}}
where $(\mathcal{M}^2_Z)^{\mu\nu}$ and $(\mathcal{M}^2_W)^{\mu\nu}$ can be found from (\ref{eq:Mg}) and $(a_{2,g})_\alpha^{~\,\beta}$ from (\ref{eg:a2g}). The reason we have left in the divergent renormalization scales (\ref{mu-tilde}) in (\ref{eff_pot_gauge}) is that since the trace depends on the dimensions of spacetime, we must not choose $n=4$ before explicitly calculating it. Due to the presence of the logarithm with Lorentz indices in general the above is a fairly non-trivial expression. However, when one limits to the case with only diagonal elements in $R^{\mu\nu}$ such as FLRW the sums can be explicitly performed. For example for the piece coming from $Z_\mu^0$ in the FLRW case we can then write
\ea{V^{(1)}_{{\rm SM}}(\varphi)_{gauge,Z}&=\f{1}{64\pi^2}\bigg[(\mathcal{M}^2_Z)_0^{\,~0}(\mathcal{M}^2_Z)_0^{\,~0}\left(\log\f{|(\mathcal{M}^2_Z)_0^{~0}|}{\mu^2}
 - \f{3}{2} \right)\,+2(a_{2,g})_0^{\,~0}\log\f{|(\mathcal{M}^2_Z\,)_0^{~0}|}{\mu^2}
 \bigg]\nonumber \\&+\f{3}{64\pi^2}\bigg[(\mathcal{M}^2_Z\,)_i^{\,~i}(\mathcal{M}^2_Z\,)_i^{\,~i}\,\left(\log\f{|(\mathcal{M}^2_Z\,)_i^{~i}|}{\mu^2}
 - \f{5}{6} \right)\,+2(a_{2,g})_i^{\,~i}\log\f{|(\mathcal{M}^2_Z\,)_i^{~i}|}{\mu^2}
 \bigg]\,,\label{eq:zed}}
where very importantly there is no sum over the repeated spatial indices denoted with "$i$". As we did for scalars and fermions, also for the gauge fields our renormalization prescription is such that the result coincides with the standard parametrization in the flat space limit (see e.g. \cite{Casas:1994qy})\footnote{The factor of $5/6$ in the second line of (\ref{eq:zed}) is the result of the interplay between the pole $\propto (4-n)^{-1}$ and the $(n-1)$-contribution from the trace.}.
The remaining $W_\nu^\pm$ piece may be obtained in a similar fashion.

With (\ref{eff_pot_scalar}),  (\ref{eff_pot_fermion}) and (\ref{eff_pot_gauge}) we have shown the calculation for the full 1-loop result including all contributions contained in the Lagrangian given in (\ref{eq:lag}). As is apparent from (\ref{eff_pot_scalar}),  (\ref{eff_pot_fermion}) and (\ref{eff_pot_gauge}) the generalization to include the complete SM is straightforward as is adding degrees of freedom beyond the SM. For an explicit example, see section \ref{sec:case}.
\subsection{The $\beta$-functions}
\label{eq:gravb} 
By following the procedures we introduced for the simple scalar field model in section \ref{sec:RGC} we can now 
derive all the $\beta$-functions of the SM in curved spacetime. This includes the well-known $\beta$-functions that can be calculated in flat space and can be found from standard references, see for example \cite{Casas:1994qy,Degrassi:2012ry,Ford:1992mv,Buttazzo:2013uya}, and the ones connected to the dynamics of a curved spacetime. The $\beta$s that are relevant in curved spacetime are defined by the operators in the purely gravitational part of the action (\ref{eq:treecurve}) and the non-minimal term proportional to $\xi$. 

With the help of formulae from section \ref{eq:appa} we can use the 1-loop approximation to the Callan-Symanzik equation (\ref{eq:CS1}) and the SM anomalous dimension \cite{DiLuzio:2014bua}
\ee{\gamma=\f{1}{16\pi^2}\bigg[
Y_2-\frac{9 g^2}{4}-\frac{3(g')^2}{4}-\zeta_W\frac{ g^2}{2}-\zeta_Z\frac{1}{4}\left(g^2+(g')^2\right)\bigg]\,,\label{eq:gamma}}

to write the gravitational $\beta$-functions

\ea{16\pi^2 \beta_{\xi} &= \label{eq:gravbetas}\bigg(\xi - \f{1}{6}\bigg)\bigg[12 \lambda +
2Y_2-\frac{3 (g')^2}{2}-\frac{9 g^2}{2}\bigg] \\ {16\pi^2}\beta_{V_\Lambda}&={2m^4}\, \\{16\pi^2}\beta_\kappa&={4}{m^2\bigg(\xi - \f{1}{6}\bigg)}\,, \\ {16\pi^2}\beta_{\alpha_1}&=2 \xi ^2-\frac{2 \xi }{3}-\frac{277}{144}\,, \label{eq:gravbetas1}\\16\pi^2 \beta_{\alpha_2}&=\frac{571}{90}\,,\label{eq:gravbetas2} \\16\pi^2 \beta_{\alpha_3}&=-\frac{293}{720}\,,\label{eq:gravbetas3}}
with
\begin{align}
Y_2 &\equiv 3(y_u^2 + y_c^2 + y_t^2) + 3(y_d^2 + y_s^2 + y_b^2) + (y_e^2 + y_{\mu}^2 + y_{\tau}^2),\nonumber\\
Y_4 &\equiv 3(y_u^4 + y_c^4 + y_t^4) + 3(y_d^4 + y_s^4 + y_b^4) + (y_e^4 + y_{\mu}^4 + y_{\tau}^4).
\end{align}
We emphasize that (\ref{eq:gravbetas} -- \ref{eq:gravbetas3}) include the contributions from the entire SM and are exhaustive in terms of the generated operators.\footnote{The particle content of the SM may be found from tables \ref{tab:contributions} and \ref{tab:contributions2}.}

The other one-loop SM $\beta$-functions are given by, for example \cite{Ford:1992mv,Buttazzo:2013uya}. Since there are curvature dependent loop corrections for all the fermions in the theory, it is not necessarily correct to ignore the light fermions, so we include the running of all the Yukawa couplings. These can be found in \cite{Luo:2002ey,Machacek:1983tz} for example,
\begin{align}
{16\pi^2}\beta_{y_t} &= {y_t}\left[\frac{3}{2}(y_t^2 - y_b^2) + Y_2 - \left(\frac{17}{12}(g')^2 + \frac{9}{4}g^2 + 8g_3^2\right) \right]\label{eq:top},\\
{16\pi^2}\beta_{y_b} &= {y_b}\left[\frac{3}{2}(y_b^2 - y_t^2) + Y_2 - \left(\frac{5}{12}(g')^2 + \frac{9}{4}g^2 + 8g_3^2\right) \right]\label{eq:bot},\\
{16\pi^2}\beta_{y_l} &= {y_l}\left[\frac{3}{2}y_l^2 + Y_2 - \left(\frac{45}{12}(g')^2 + \frac{9}{4}g^2\right) \right]\label{eq:lepton},\\
{16\pi^2}\beta_{\lambda} &=24\lambda^2 - 3\lambda \left((g')^2 + 3g^2\right) + \f{3}{4}\left(\frac{1}{2}(g')^4 + (g')^2g^2 + \frac{3}{2}g^4\right) + 4Y_2\lambda - 2Y_4,\\
{16\pi^2}\beta_{m^2} &={m^2}\left[12\lambda - \frac{3}{2}(g')^2 - \frac{9}{2}g^2 + 2Y_2\right],\\
{16\pi^2}\beta_{g'} &= \frac{41}{6}{(g')^3} ,\qquad
{16\pi^2}\beta_{g} =  -\frac{19}{6}{g^3},\qquad
{16\pi^2}\beta_{g_3} = -7 {g_3^4},\nonumber\\
\end{align}
where $\beta_{y_l}$ is the lepton beta function for $l = e,\mu,\tau$. For reference, the beta functions are defined as $\beta_{X} \equiv \mu(\dd X/\dd \mu)$, $g'$ is the $U(1)$ coupling, 
$g$ is the SU(2) coupling, and $g_3$ the $SU(3)$ coupling. Beta functions for the other generations of fermions can be obtained from eqs.~(\ref{eq:top}) and (\ref{eq:bot}) by substituting $y_t \rightarrow y_u, y_c$ and $y_b\rightarrow y_d,y_s$, leaving the gauge couplings and $Y_2$ the same.

In priciple also the gauge parameters $\zeta_Z$ and $\zeta_W$ would run according to their respective $\beta$-functions. Their running is however not relevant for a one loop calculation: the $\zeta$'s only enter in the loop correcting and do not couple to other one loop $\beta$-functions making the running a two loop effect and with no loss of generality one may treat them as constant, which will be our choice.

For the boundary conditions of the running couplings at the EW scale $t=0$, we use the precise matching relationships between pole masses and $\overline{\rm MS}$ parameters found in \cite{Buttazzo:2013uya}, supplemented with one-loop results for the remaining fermions in the theory \cite{PhysRevD.51.1386}. Unless otherwise stated, we used top quark and Higgs boson pole masses of $M_t = 173.34 \rm{\,GeV}$ and $M_h = 125.15\rm{\,GeV}$ respectively, with other pole masses found in \cite{pdg}.

\section{A case study: the SM in de Sitter space}
\label{sec:case}

The general results derived above using the heat kernel method hold for an arbitrary curved spacetime.  As a specific example, we apply them in the de Sitter space where 
\ee{R=12H^2\,,\quad R_{\mu\nu}R^{\mu\nu}=36H^4\,,\quad R_{\mu\nu\delta\eta}R^{\mu\nu\delta\eta}=24H^4\,,
\label{dSRiemann}} 
and the Hubble rate $H$ is constant. 

Substituting these into the expressions derived in Sections 3 and 4,  we find that the 1-loop contribution to the effective potential of the SM Higgs in the $\overline{\rm MS}$ scheme, including the complete set of quarks and leptons as well as the photon and the gluons, is given by 
\ea{
	V_{\rm SM}^{(1)}(\varphi)& = 
	\f{1}{64\pi^2} \sum\limits_{i=1}^{31}\bigg\{ n_i\mathcal{M}_i^4\bigg[\log\left(\f{|\mathcal{M}_i^2|}{\mu^2}\right) - d_i \bigg] +{n'_i}H^4\log\left(\f{|\mathcal{M}_i^2|}{\mu^2}\right)\bigg\}\,,
	\label{potential}}
where the inputs can be read from tables \ref{tab:contributions} and \ref{tab:contributions2}. They are split as the degrees of freedom that directly couple to the Higgs in table \ref{tab:contributions} and degrees of freedom that do not in table \ref{tab:contributions2}. Recall that our computation gives the UV limit of the effective potential.  
\begin{table}
	\caption{\label{tab:contributions}Contributions to the effective potential (\ref{potential}) with tree-level couplings to the Higgs. $\Psi$ stands for $W^{\pm}$, $Z^0$, the 6 quarks q, the 3 charged leptons $l$, the Higgs $h$, the Goldstone bosons $\chi_{W}$ and $\chi_{Z}$ and the ghosts $c_{W}$ and $c_{Z}$. The masses are defined as in (\ref{eq:masses}).}
	\vspace{2mm}
	\begin{center}
		{\tabulinesep=1.6mm
			\begin{tabu}{|c||ccccc|}
				\hline
				$\Psi$ & $~~i$ & $~~n_i$   & $~d_i$    &$~n'_i$      &$\hspace{-2.6cm}\mathcal{M}_i^2$    \\\hhline{|=#=====|}
				$~$ & $~~1$  & $~~2$       & $\quad{3}/{2}~~$        & $-34/15$        &  $\hspace{-2.8cm}m^2_W+H^2$      \\
				$~W^\pm$ & $~~2$  & $~~6$       & $\quad{5}/{6}~~$       & $-34/5$        &  $\hspace{-2.8cm}m^2_W+H^2$       \\
				$~$ & $~~3$  & $-2$      & $\quad{3}/{2}~~$         & $~~4/15$        & $\hspace{-2.7cm} ~m^2_W-2H^2$        \\\hline
				$~$ & $~~4$  & $~~1$        & $\quad{3}/{2}~~$ & $-17/15$        & $\hspace{-2.8cm} m^2_Z+H^2$      \\
				$Z^0$ & $~~5$  & $~~3$        & $\quad{5}/{6}~~$ & $-17/5$        & $\hspace{-2.8cm} m^2_Z+H^2$     \\
				$~$ & $~~6$  & $-1$      & $\quad{3}/{2}~~$ & $~~2/15$        & $\hspace{-2.7cm} ~m^2_Z-2H^2$     \\\hline
				q & $7-12$  & $-12$     & $\quad{3}/{2}~~$    & $~~38/5$        & $\hspace{-2.8cm} m^2_{q}+H^2$      \\\hline
				$l$ & $13-15$  & $-4$     & $\quad{3}/{2}~~$    & $~38/15$        & $\hspace{-2.8cm} m^2_{l}+H^2$      \\\hline
				$h$ & $~16$  & $~~1$       & $\quad{3}/{2}~~$          & $-2/15$      & $\hspace{-0.9cm}m_h^2+12(\xi-{1}/{6})H^2$  \\\hline
				${\chi}_W$ & $~17$  & $~~2$      & $\quad{3}/{2}~~$           & $-4/15$      & $\,\,\,~~\quad m_\chi^2\,+\zeta_W m^2_W+12(\xi-{1}/{6})H^2$   \\\hline
				${\chi}_Z$ & $~18$  & $~~1$       & $\quad{3}/{2}~~$           & $-2/15$      & $\,~~\quad m_\chi^2+\zeta_Z m^2_Z\,+12(\xi-{1}/{6})H^2$   
				\\\hline ${c}_W$ & $~19$  & $-2$       & $\quad{3}/{2}~~$           & $~~4/15$      & $\hspace{-1.9cm}\zeta_W m^2_W-2H^2$  
				\\\hline
				${c}_Z$ & $~20$  & $-1$      & $\quad{3}/{2}~~$            & $~~2/15$      & $\hspace{-2.1cm}\zeta_Z m^2_Z-2H^2$   
				\\\hline
			\end{tabu}
		}
	\end{center}
	
\end{table}
\newpage
\begin{table}
	\caption{\label{tab:contributions2}Contributions to the effective potential (\ref{potential}) that do not to couple to the Higgs at tree-level. $\Psi$ stands for the photon $\gamma$, the 8 gluons $g$, the 3 neutrinos $\nu$ and the ghosts $c_\gamma$ and $c_g$. }
	\vspace{2mm}
	\begin{center}
		{\tabulinesep=1.6mm
			\begin{tabu}{|c||ccccc|}
				\hline
				$\Psi$ & $~~i$ & $~~n_i$   & $~d_i$    &$~n'_i$      &$\mathcal{M}_i^2$    \\\hhline{|=#=====|}
				$~$ & $~21$  & $~~1$       & $\quad{3}/{2}~~$        & $-17/15$        &  $H^2$      \\
				$~\gamma$ & $~22$  & $~~3$       & $\quad{5}/{6}~~$       & $-17/5$        &  $H^2$       \\
				$~$ & $~23$  & $-1$      & $\quad{3}/{2}~~$         & $~~2/15$        & $ -2H^2$        \\\hline
				$~$ & $~24$  & $~~8$        & $\quad{3}/{2}~~$ & $-136/15$        & $ H^2$      \\
				$g$ & $~25$  & $~~24$        & $\quad{5}/{6}~~$ & $-136/5$        & $ H^2$     \\
				$~$ & $~26$  & $-8$      & $\quad{3}/{2}~~$ & $~~16/15$        & $-2H^2$     \\\hline
				$\nu$ & $27-29$  & $-2$      & $\quad{3}/{2}~~$           & $~~19/15$      & $H^2$   \\\hline
				${c}_\gamma$ & $~30$  & $-1$       & $\quad{3}/{2}~~$           & $~~2/15$      & $-2H^2$  
				\\\hline
				${c}_g$ & $~31$  & $-8$      & $\quad{3}/{2}~~$            & $~~16/15$      & $-2H^2$   
				\\\hline
			\end{tabu}
		}
	\end{center}
	
\end{table}

A generic expression for $M_i^2$ is thus a function of the form $M_i^2(\varphi_{\rm{cl}},\mu) = \kappa_i(\mu)\frac{Z(M_t)}{Z(\mu)}\varphi_{\rm{cl}}^2 - \kappa'_i(\mu) + \theta_i(\mu)H^2$, where the coupling-dependent $\kappa_i,\kappa'_i,\theta_i$ can be read from tables \ref{tab:contributions} and \ref{tab:contributions2}, using field-dependent masses for all the Standard Model particles. The one-loop effective potential with running couplings is then given by 
\baq
\nonumber
V^{\rm eff}_{\rm SM}(\varphi(\mu))&=&-\f{1}{2}m^2(\mu)\varphi^2(\mu)+\f{\xi(\mu)}{2}R\varphi^2(\mu)+\f{\lambda(\mu)}{4}\varphi^4(\mu)+V_\Lambda(\mu)-12\kappa(\mu) H^2+\alpha(\mu) H^4\\
&&+ \f{1}{64\pi^2} \sum\limits_{i=1}^{31}\bigg\{ n_i\mathcal{M}_i^4(\mu)\bigg[\log\left(\f{|\mathcal{M}_i^2(\mu)|}{\mu^2}\right) - d_i \bigg] +{n'_i}H^4\log\left(\f{|\mathcal{M}_i^2(\mu)|}{\mu^2}\right)\bigg\}\,.
\label{eq:VeffSMdS}
\eaq
The different gravitational terms $\alpha_{1} R^2$, $\alpha_2 R_{\mu\nu}R^{\mu\nu}$, $\alpha_3 R_{\mu\nu\delta\eta}R^{\mu\nu\delta\eta}$ in (\ref{eq:treecurve}) have combined to a single term $\alpha(\mu) H^4$ due to the de Sitter relations (\ref{dSRiemann}). The $\beta$ function of the coupling $\alpha$ is determined by using eqs. (\ref{eq:gravbetas1} -- \ref{eq:gravbetas3}) which yield 
\ee{\beta_{\alpha}=\f{1}{16\pi^2}\bigg[288 \xi ^2-96 \xi -\frac{1751}{30}\bigg] \label{eq:betaC2}\,.}
All other $\beta$ functions are directly given in section \ref{eq:gravb}. 

The renormalization group improved potential is found by numerically solving for the full set of $\beta$ functions and substituting the results into (\ref{eq:treecurve}).  
We use a modification of a method recently employed by two of the authors in \cite{vacstab17,PhysRevD.97.025012}. Briefly, this method consists of computing the running of the couplings at a set of discrete points, and using these to construct a $C_1$ continuous interpolating piecewise polynomial to describe the running in logarithmic space. This results in function $g_i(\mu)$ for the couplings, and together with a scale choice $\mu(\varphi)$, gives a numerical expression for the potential. The final expression is then:
\begin{align}
V_{SM}^{RGI}(\varphi_{\rm{cl}}) =& \frac{1}{2}\bigg[-m^2(\mu_{*}(\varphi_{\rm{cl}})) + \xi(\mu_{*}(\varphi_{\rm{cl}}))R\bigg]\frac{Z(M_t)}{Z(\mu_{*}(\varphi_{\rm{cl}}))}\varphi_{\rm{cl}}^2 + \frac{\lambda(\mu_{*}(\varphi_{\rm{cl}}))}{4}\frac{Z^2(M_t)}{Z^2(\mu_{*}(\varphi_{\rm{cl}}))}\varphi_{\rm{cl}}^4\nonumber\\
& + V_{\Lambda}(\mu_{*}(\varphi_{\rm{cl}})) - 12\kappa(\mu_{*}(\varphi_{\rm{cl}}))H^2 + \alpha(\mu_{*}(\varphi_{\rm{cl}}))H^4 \nonumber\\
& + \frac{1}{64\pi^2} \sum\limits_{i=1}^{31}\bigg\{ n_i\mathcal{M}_i^4(\varphi_{\rm{cl}})\bigg[\log\left(\frac{|\mathcal{M}_i^2(\varphi_{\rm{cl}})|}{\mu_{*}^2(\varphi_{\rm{cl}})}\right) - d_i \bigg] +{n'_i}H^4\log\left(\f{|\mathcal{M}_i^2(\varphi_{\rm{cl}})|}{\mu_{*}^2(\varphi_{\rm{cl}})}\right)\bigg\}\,,
\end{align}
where $\mathcal{M}_i^2(\varphi_{\rm{cl}}) = \mathcal{M}_i(\varphi_{\rm{cl}},\mu_{*}(\varphi_{\rm{cl}}))$ is the relevant mass-term defined at scale $\mu_{*}(\varphi_{\rm{cl}})$, $\varphi_{\rm{cl}} = \varphi(M_t)$ is the field evaluated at the electroweak scale, and $\alpha = 144\alpha_1 + 36\alpha_2 + 24\alpha_3$.

\subsection{Scale Choice}
\label{sec:scaleChoice}

The renormalisation group improvement procedure discussed in section~\ref{sec:RGC} consists of choosing the renormalisation
scale in such a way that the loop correction vanishes. Applying the same method to the full SM case amounts to setting $\mu=\mu_*$ given by the  condition 
\begin{align}
\sum_{i=1}^{31}
\left[n_i{\cal M}_i^4(\mu_*,\varphi)
\left(\log\left(\frac{|{\cal M}_i^2(\mu_*,\varphi)|}{\mu_*^2}\right) - d_i\right) + n'_iH^4\log\left(\frac{|{\cal M}_i^2(\mu_*,\varphi)|}{\mu_*^2}\right)\right] = 0\label{eq:muvanish}
\end{align}

However, unlike in the simple scalar theory of section~\ref{sec:RGC}, this equation may now have several solutions as it contains multiple terms with different mass scales ${\cal M}_i$. 
This is illustrated in fig. \ref{fig:fmuphi} which shows two solutions of eq. ({\ref{eq:muvanish}).
The individual logarithms ${\cal M}_i^4\; {\rm log}({\cal M}_i^2/\mu^2))$ are not necessarily small for all solutions. Therefore, eq. ({\ref{eq:muvanish}) alone is not enough to ensure convergence of the one-loop effective potential which is accurate up to log squared corrections, such as ${\cal O}((y_t/16\pi^2)^2 {\cal M}_t^4 {\rm log}^2({\cal M}_t^2/\mu^2))$. 
Fig. \ref{fig:fmuphi} also shows the range where the dominant log contributions are small for each solution. Outside this range the logs grow large and using $\mu_*$ given by eq. ({\ref{eq:muvanish}) does not give a reliable result. This is demonstrated in fig. \ref{fig:twosolpots} which shows the effective potentials computed for the two different solutions of eq. ({\ref{eq:muvanish}).  

\begin{figure}[htb]
\centering
\includegraphics[height = 0.4\textheight]{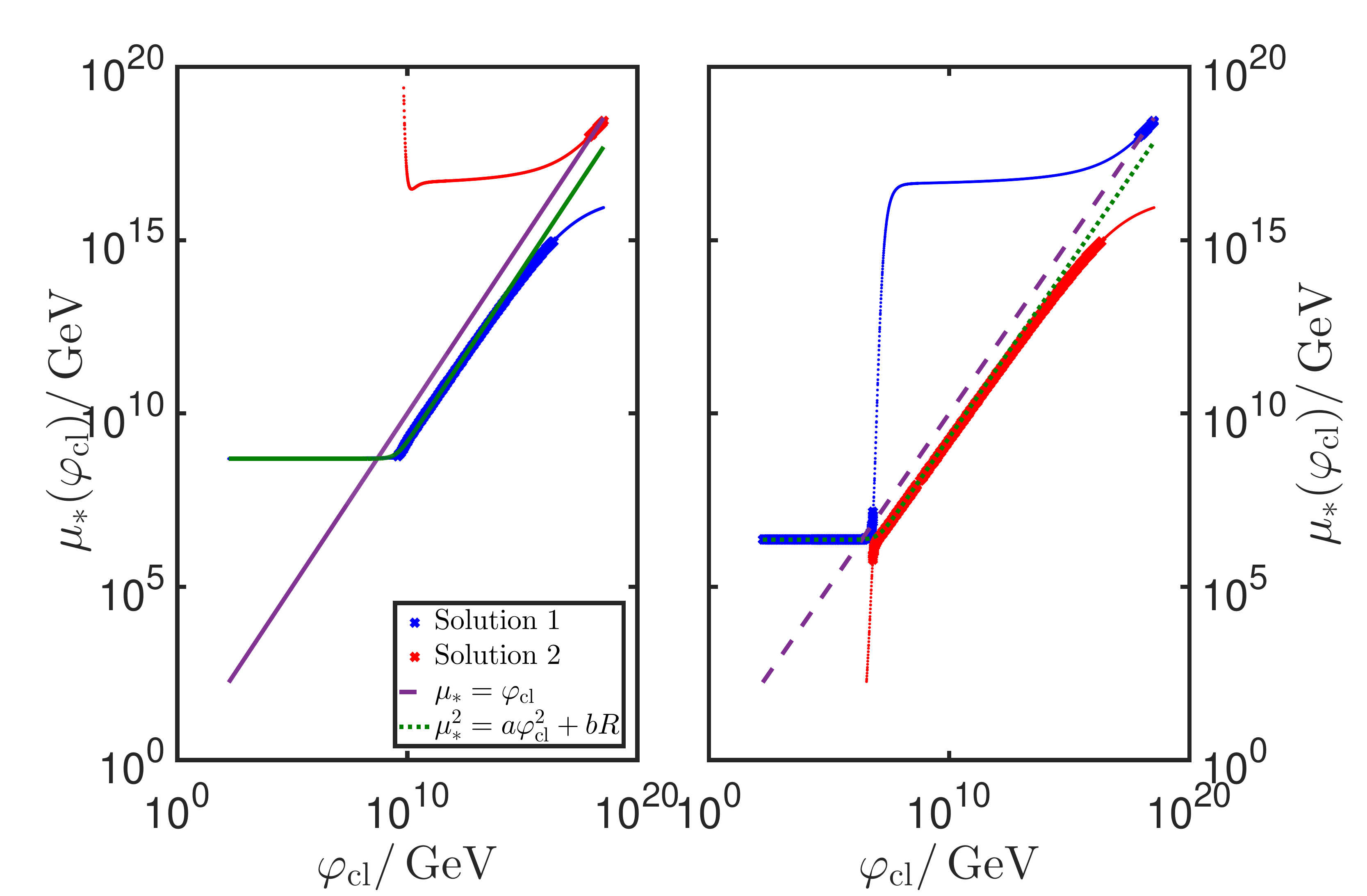}
    \caption{\label{fig:fmuphi}Two solutions of Eq.~(\ref{eq:muvanish}) 
     for a range of $\varphi_{\rm cl}$, and two different values of $\xi_{EW}$ and $H$. A `best fit' choice as in Eq. \ref{eq:phiR} is also plotted, along-side the $\mu = \varphi$ choice. Left: $\xi_{\rm{EW}} = \xi(M_t) = 1/6$, $H = 10^{9}\rm{\,GeV}$, $a = 0.0215$ and $b = 0.0203$. Right: $\xi_{\rm{EW}} = \xi(M_t) = -1$, $H = 10^{6}\rm{\,GeV}$, $a = 0.0414$, $b = 0.4332$. Large crosses denote points where the log contributions of the dominant fields, $W,Z,\varphi,t$, all satisfy $\left|\log\frac{M^2}{\mu^2}\right| < 5$, small dots where this is not true.}
\end{figure}

We could accompany eq. (\ref{eq:muvanish}) with the additional condition that the individual logarithms ${\cal M}_i^4\; {\rm log}({\cal M}_i^2/\mu^2))$ remain small. However, there are two problems with this procedure. First, it is not always possible to find any solution $\mu_*$ for which the logs are small while eq. (\ref{eq:muvanish}) is satisfied exactly. Second, even when such a solution exists, it may not give a continuous potential over the whole range of $\varphi_{\rm cl}$.

To obtain a continuous potential, one has to relax the requirement that the one loop correction vanishes completely, and instead choose the renormalization
scale in a continuous way so that the loop correction remains small albeit non-zero. A simple scale choice that approximatively achieves this is
\begin{equation}
\mu^2 = a \varphi_{\rm{cl}}^2 + b R,\label{eq:phiR}
\end{equation}
where
\begin{equation}
\varphi_{\rm{cl}} = \varphi(M_t).
\end{equation}
There are several ways to determine the constants $a,b$, for example as the best fit values that interpolate between the different solutions of eq. (\ref{eq:muvanish}), or chosen to minimise the size of the logarithm terms in the potential. A similar choice without the numerical fitting was used in \cite{vacstab7,Herranen:2015ima}. The scale choice (\ref{eq:phiR}) is dominated by the curvature $R$ for small $\varphi$ and by the field for large $\varphi$ and can therefore minimize loop terms of type ${\rm log}({\cal M}^2/\mu^2)$ where ${\cal M}^2$ is a sum of $R$ and $\varphi$ dependent entries. This works well if the effective potential does not depend too strongly
on the scale $\mu$ across the relevant range.
As an example, we plot in figure \ref{fig:mudep} the effective potential $V(\varphi,\mu)$ for different $\mu$ at constant $\varphi$, $H = 10^9\rm{\,GeV}$ and $\xi = \frac{1}{6}$. If the scale is well chosen, then it should lie in a relatively flat region of the resulting curve. It is noticed that the renormalization scale can drastically change $V(\varphi,\mu)$ if $\varphi$ is close to the barrier. This is perhaps unsurprising, as the potential rapidly changes from positive to negative values there.

\begin{figure}[htb!]
	\centering
	\includegraphics[height = 0.4\textheight]{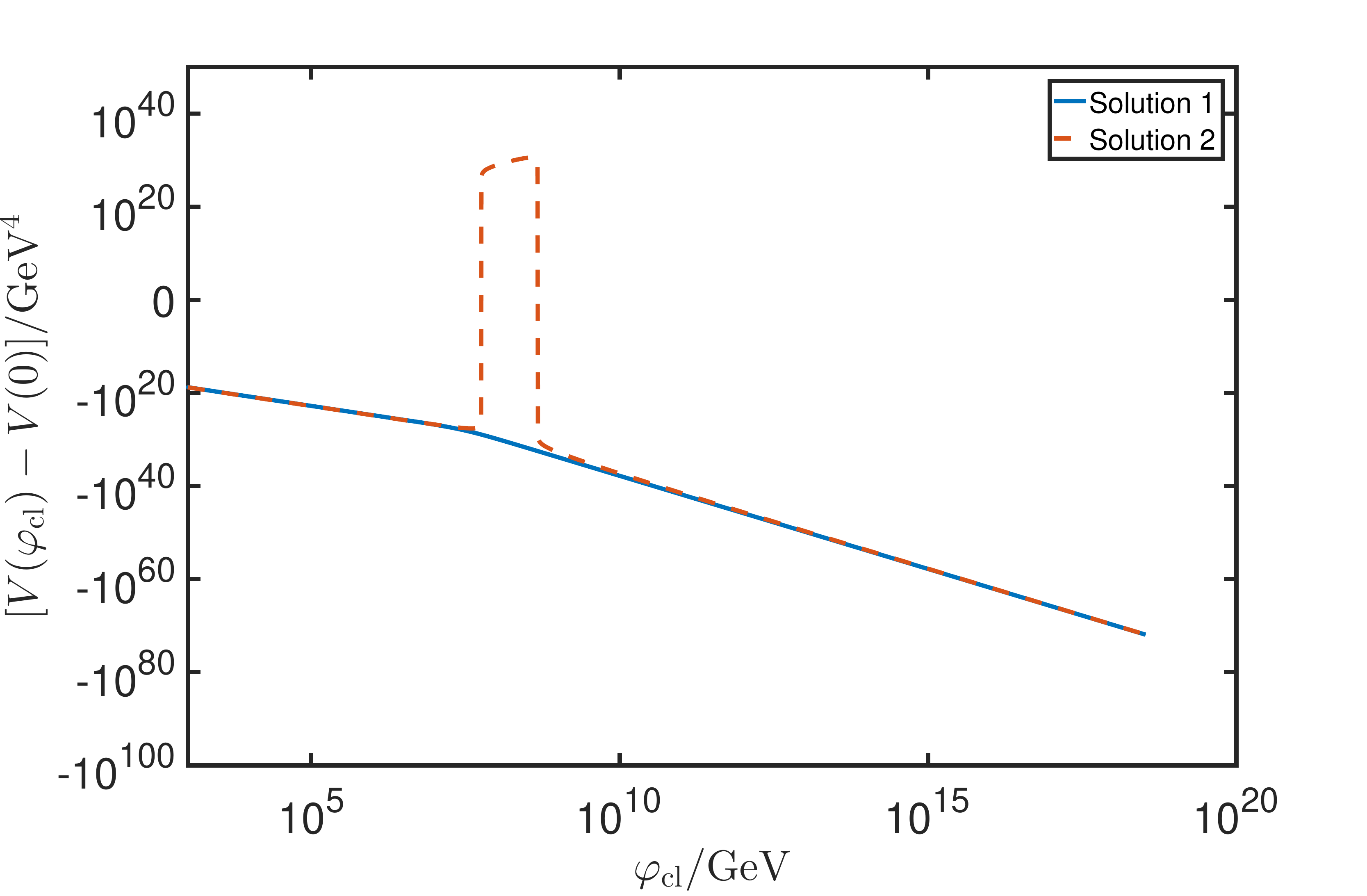}
	\caption{\label{fig:twosolpots}RGI Potentials associated to solution 1 and solution 2 for $\xi_{EW} = -1,H = 10^{6}\rm{\,GeV}$, from fig. \ref{fig:fmuphi}. Note that solution 1 is completely unstable, while solution 2 possesses a barrier. The scale $\mu_{\rm{inst}}$ is the renormalization scale at which $\lambda(\mu_{\rm{inst}}) = 0$.}
\end{figure}

\FloatBarrier

\begin{figure}[htb]
	\centering
	\includegraphics[height = 0.4\textheight]{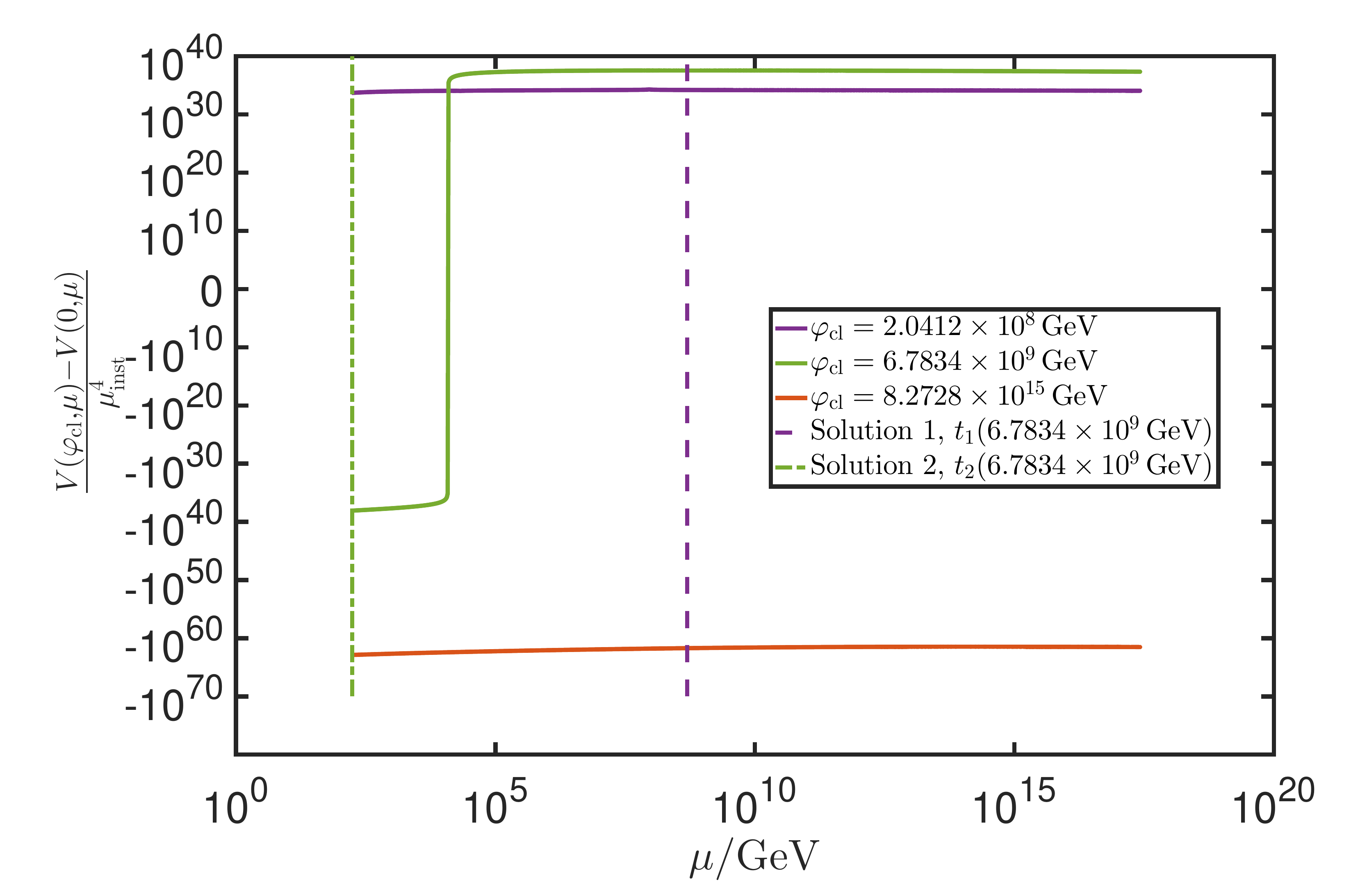}
	\caption{\label{fig:mudep}Renormalisation scale dependence of the approximated effective potential for $H = 10^9\rm{\,GeV}$ and $\xi_{\rm{EW}} = 1/6$. For most values of $\varphi_{\rm cl}$, the potential does not vary considerably. The exception is in the vicinity of the barrier, where changing $\mu$ can rapidly shift the potential from positive to negative values, since it moves around the zero of the potential.}
\end{figure}

\FloatBarrier

\subsection{Potential}
\label{sec:potential}
To consider a wide range of possible scenarios, we plot in figures \ref{fig:xim1} to \ref{fig:xi1o6} the potential as a function of the 
``classical'' field $\varphi_{\rm cl}$.
We also plot the potential in units of the instability scale $\mu_{\rm inst}$, defined as the renormalisation scale at which  $\lambda(\mu_{\rm inst})=0$, rather than in units of GeV.
The reason for this is that the instability scale given by the one-loop calculation ($\mu_{\rm inst}=9.45\times 10^{7}\rm{\,GeV}$) is quite different from the more accurate three-loop value ($\mu_{\rm inst}=6.59\times 10^{9}\rm{\,GeV}$) (computed using 3 loop running of the Standard Model beta functions, see for example \cite{Zoller:2014cka}), 
and this is reflected in the scale of the corresponding potentials. Expressing the potential in units of $\mu_{\rm inst}$ should therefore allow a more meaningful comparison and 
more reliable physical conclusions.

\begin{figure}[htb]
	\centering
	\includegraphics[height = 0.4\textheight]{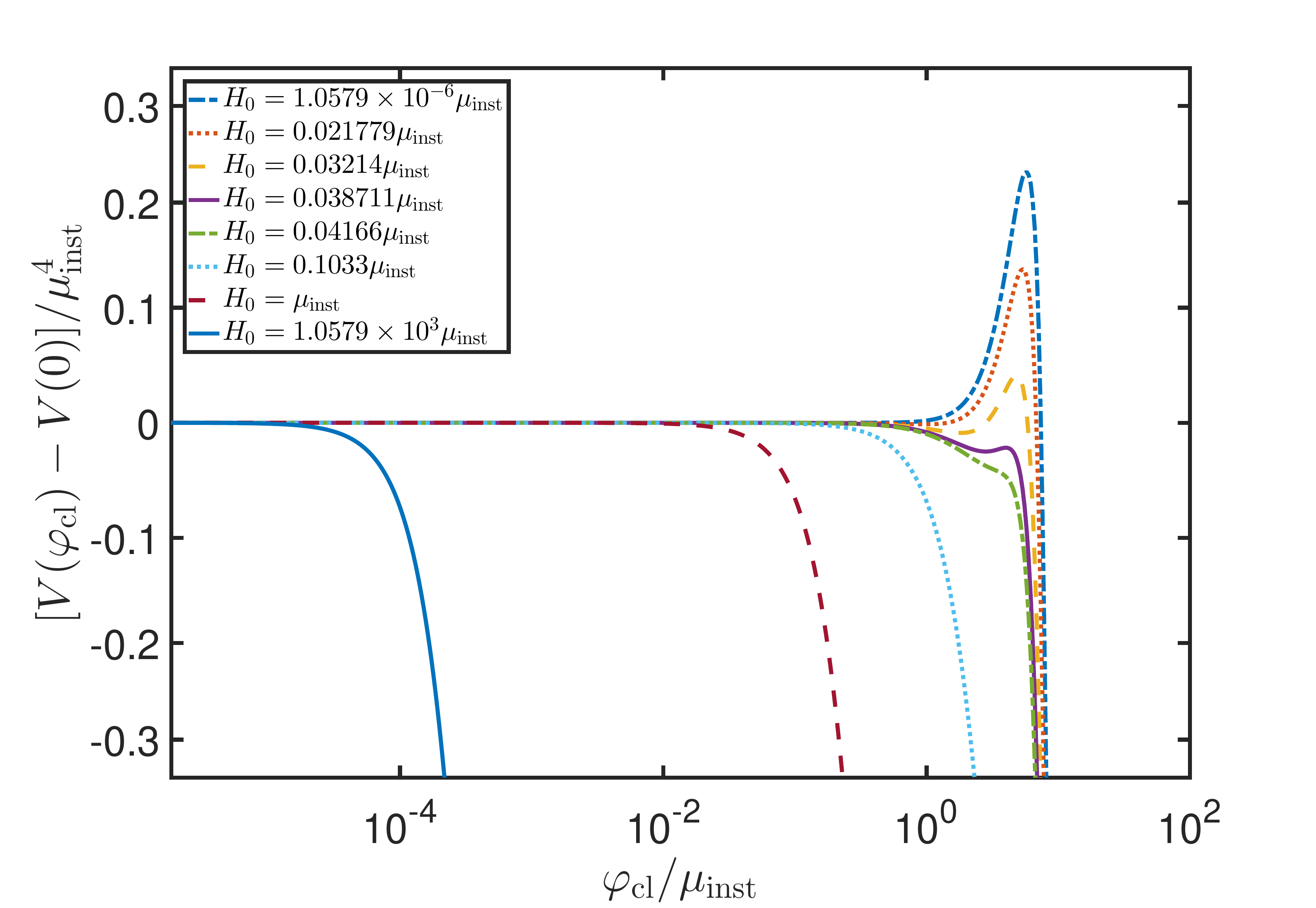}
	\caption{\label{fig:xim1}Example RGI potentials for $\xi_{\rm{EW}} = -1$, in units of $\mu_{\rm{inst}}$ defined by $\lambda(\mu_{\rm{inst}}) = 0$.}
\end{figure}
\begin{figure}[htb]
	\centering
	\includegraphics[height = 0.4\textheight]{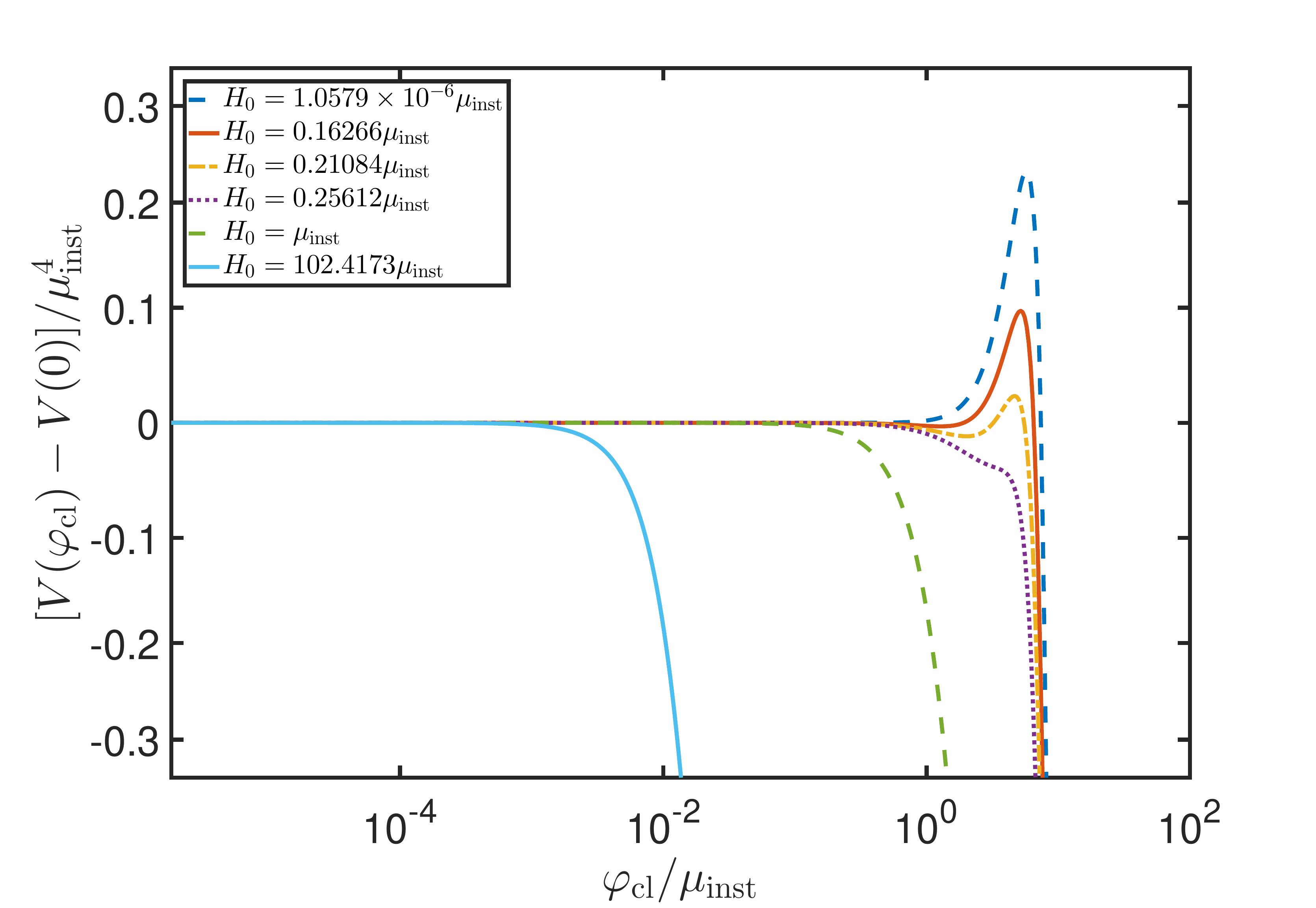}
	\caption{\label{fig:xim0}Example RGI potentials for $\xi_{\rm{EW}} = 0$, in units of $\mu_{\rm{inst}}$ defined by $\lambda(\mu_{\rm{inst}}) = 0$.}
\end{figure}
\begin{figure}[htb]
	\centering
	\includegraphics[height = 0.4\textheight]{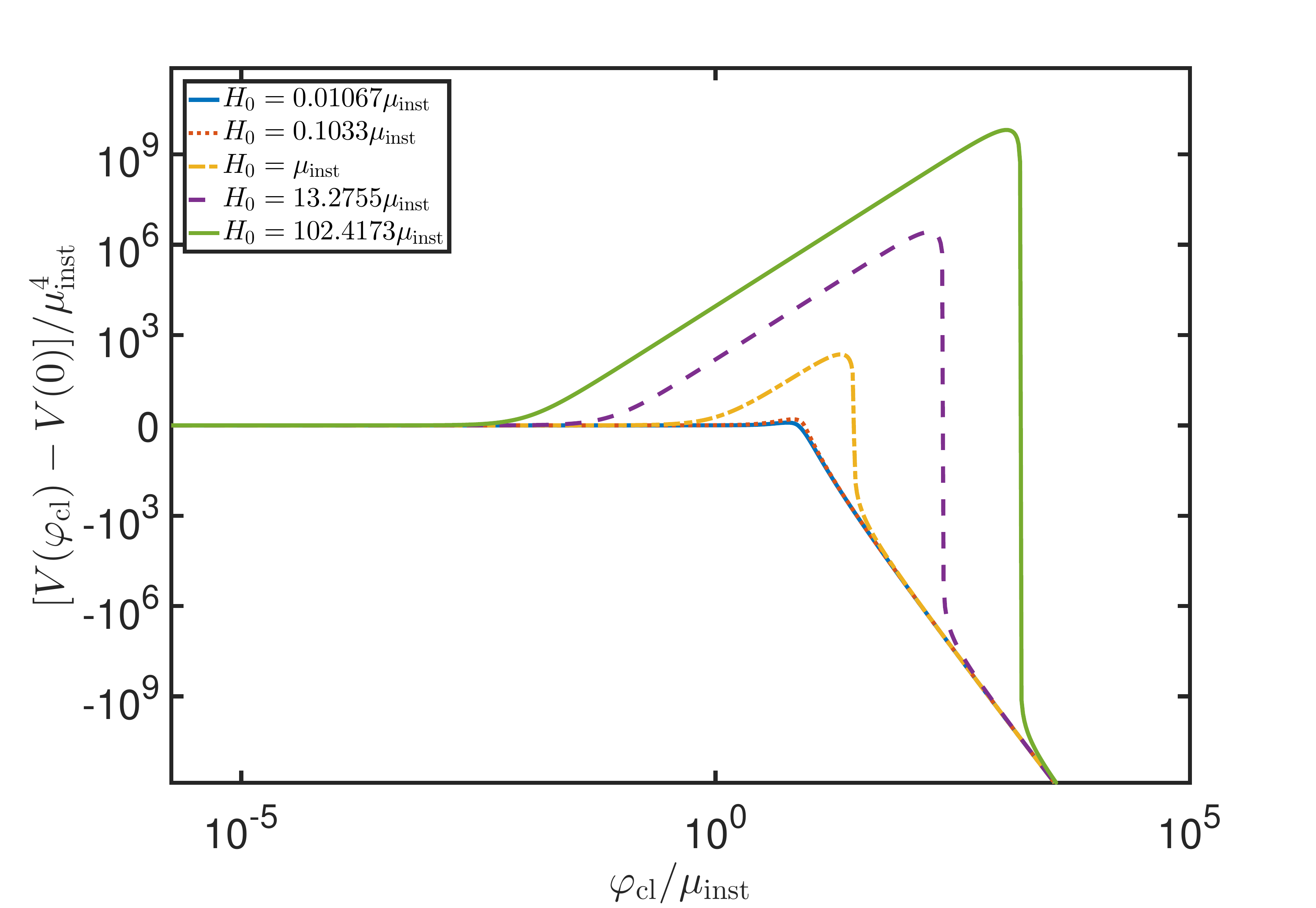}
	\caption{\label{fig:xi1o6}Example RGI potentials for $\xi_{\rm{EW}} = 1/6$, in units of $\mu_{\rm{inst}}$ defined by $\lambda(\mu_{\rm{inst}}) = 0$.}
\end{figure}
In figs.~\ref{fig:xim1},~\ref{fig:xim0},~\ref{fig:xi1o6},
we show the effective potential for a range of different Hubble rates, $H$, and non-minimal couplings $\xi_{EW} = \xi(\mu = M_t)$ (as evaluated at the electroweak scale). 
The scale $\mu$ is chosen 
to be Eq.~(\ref{eq:phiR}),
in order to give a continuous potential for all $\varphi$. There are various ways of picking the coefficients $a$ and $b$; in the case of the potentials plotted in figs. ~\ref{fig:xim1},~\ref{fig:xim0},~\ref{fig:xi1o6}, we choose $a$ and $b$ such that a weighted sum of the squares of the logs, 
\begin{equation}
S = \frac{\sum_{i}{\cal M}_i^4\log ({\cal M}_i^2/\mu^2)^2}{\sum_i{\cal M}_i^4},
\end{equation}
is minimized at both a small scale ($\varphi_{\rm{cl}} = M_t$) and a large scale ($\varphi_{\rm{cl}} = M_{\rm{P}}$).

Notice that for negative $\xi$, sufficiently large $H$ erases the barrier altogether. For positive $\xi$, the opposite occurs: larger $H$ raises the height of the barrier. The case of $\xi = 0$ at the electroweak scale is effectively the same as a negative $\xi$ scenario, because the coupling runs to negative values at scales above the electroweak scale, and the potential is evaluated at $\mu^2 = bR$ for small $\varphi_{\rm cl}$ (thus, unless $R = 0$, the relevant scale is never the electroweak scale for any $\varphi_{\rm cl}$, and so $\xi < 0$ everywhere). 

\FloatBarrier
\subsection{Gauge Dependence}
\label{sec:gauge}
The gauge dependence of the instability scale for the SM in flat space was recently investigated in \cite{DiLuzio:2014bua,Andreassen:2014gha,Espinosa:2016nld,Espinosa:2016uaw}.

It is well known that the stationary points of the effective potential for a gauge theory are independent of the gauge, even if the potential is not. This is described by the Nielsen identity for the effective potential \cite{Nielsen:1975fs}
\ee{\label{eq:nielsen}\f{\partial V^{\rm SM}_{\rm eff}(\varphi)}{\partial\zeta}=C(\varphi,\zeta)\f{\partial V^{\rm SM}_{\rm eff}(\varphi)}{\partial\varphi}\,.}
From the above it trivially follows that at extremal points the effective potential is gauge independent. In flat space this fact ensures that vacuum decay rates are gauge independent quantities \cite{Plascencia:2015pga} and the same can be seen for the Hawking-Moss instanton in de Sitter space, since it's action depends only on the height of the barrier (see section \ref{sec:stability}). 

The Nielsen identity, however, only applies to the RGI effective potential up to the order of truncation, and therefore
it is important to check how much its extremal points depend on the gauge choice. 
In fig. \ref{fig:gauge} we plot the potential for different choices of the gauge parameter. 
\begin{figure}[htb]
	\centering
	\includegraphics[height = 0.4\textheight]{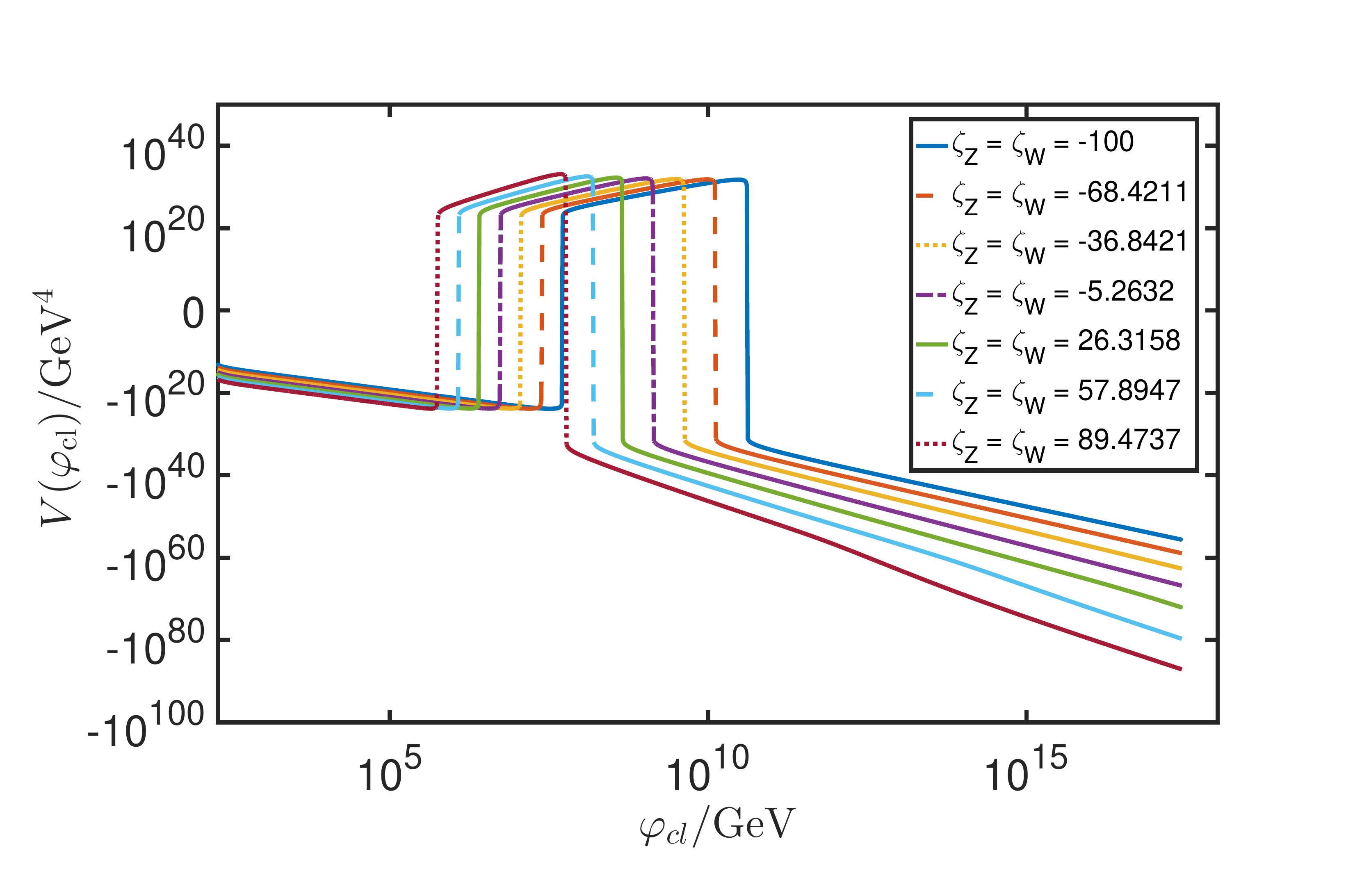}
	\caption{\label{fig:gauge}Gauge dependence of the potential for $H= 10^6\rm{\,GeV}$ and $\xi_{\rm{EW}} = 0$. The extremal should be gauge-independent and our one-loop approximation satisfies this condition relatively well. Y-axis is scaled as $y = \rm{sign}(x)\log(1 + |x|)$, which is logarithmic for large positive and negative values - this causes the sharp apparent change across zero.}
\end{figure}
This shows that across a reasonably large parameter space for the gauge parameters $\zeta_W, \zeta_Z$, the height of the potential barrier stays roughly the same (outside this parameter range, a numerical instability related to choosing the scale that sets the sum of the quantum corrections was encountered). Fig. \ref{fig:gauge}, computed with the `physical' (smallest log) solution, also shows that unlike the extrema, the scale at which the potential turns over to negative values is sensitive to the gauge parameter, due in most part to the gauge dependence of the anomalous dimension.
\FloatBarrier

\subsection{Implications for Vacuum Stability\label{sec:stability}}
Stability of the vacuum during inflation depends on two factors: the nucleation of true vacuum bubbles, and the expansion of space-time. 
This issue has been considered, for example, by \cite{vacstab14,vacstab13,East:2016anr}. Here, we will present a simplified analysis based on the Hawking-Moss instanton. This neglects some potentially important effects: see \cite{vacstab13} for a discussion of this. Note also that below a certain critical Hubble rate, Coleman de Luccia (CdL) bounces are expected to dominate~\cite{PhysRevD.21.3305,PhysRevD.69.063518}.\\
The decay rate predicted by Hawking-Moss instantons is of the form
\begin{equation}
\Gamma = c H^4 \exp(-B_{\rm{HM}}),
\end{equation}
where $c$ is an $O(1)$ constant, and the prefactor is proportional to $H^4$ on dimensional grounds\footnote{A proper analysis would involve computing the full functional determinant\cite{PhysRevD.21.3305}; the precise pre-factor, however, will not matter much due to the exponential dependence on $B_{\rm{HM}}$}. The decay exponent, $B_{\rm{HM}}$, is given by\cite{Hawking:1981fz}, including non-minimal coupling, by
\begin{equation}
B_{\rm{HM}} = 24\pi^2M_{\rm{P}}^4\left[\frac{1}{V_0(\varphi_{\rm{fv}})}\left(1 - \frac{\xi\varphi_{\rm{fv}}^2}{M_{\rm{P}}^2}\right)^2-\frac{1}{V_0(\varphi_{\rm{HM}})}\left(1 - \frac{\xi\varphi_{\rm{HM}}^2}{M_{\rm{P}}^2}\right)^2\right].\label{eq:fullHMB}
\end{equation}
Here $\varphi_{\rm{fv}}$ is the false vacuum field value, $\varphi_{\rm{HM}}$ the top of the barrier, and $V_0$ represents the effective potential \emph{without} the $\frac{1}{2}\xi\varphi^2R$ non-minimal coupling term. This result includes the effect of gravitational backreaction of the Hawking-Moss solution, and can be seen most easily in the Einstein frame, where the potential takes the form\cite{Bezrukov:2007ep}
\begin{equation}
\tilde{V}(\tilde{\varphi}) = \frac{V_0(\varphi(\tilde{\varphi}))}{\left(1 - \frac{\xi\varphi^2(\tilde{\varphi})}{M_{\rm{P}}^2}\right)^2}.
\end{equation}
$\tilde{V}$ and $\tilde{\varphi}$ are the canonical Einstein frame potential and field, respectively. Thus in the Einstein frame, Eq.~(\ref{eq:fullHMB}) is just the usual Hawking-Moss formula. Generically, the action of an $O(4)$ symmetric bounce solution is given by
\begin{equation}
S = 2\pi^2\int_{0}^{r_{\rm{max}}}\dd r a^3(r)\left[\frac{1}{2}\dot{\varphi}^2 + V_0(\varphi) - \frac{M_{\rm{P}}^2}{2}\left(1 - \frac{\xi\varphi^2}{M_{\rm{P}}^2}\right)R\right],\label{eq:action}
\end{equation}
where $a(r)$ is the scale factor in the Euclidean metric, $\dd s^2 = \dd r^2 + a^2(r)\dd\Omega_3^2$ and $r$ is a (Euclidean) radial co-ordinate. Using Eq. (\ref{eq:action}) we can compute $B_{\rm{HM}}$ of the Hawking-Moss instanton in the limit where $R$ is fixed - the so called `fixed background approximation'. This is given by
\begin{equation}
B_{\rm{HM}} = \frac{8\pi^2\Delta V_{\xi}(\varphi_{\rm{HM}})}{3H^4},
\end{equation}
where $H$ is the Hubble rate, $R = 12H^2$, and 
\begin{equation}
\Delta V_{\xi}(\varphi) = V_0(\varphi) + \frac{\xi}{2}\varphi^2R - V_0(\varphi_{\rm{fv}}) - \frac{\xi}{2}\varphi_{\rm{fv}}^2.
\end{equation}
This means that $B_{\rm{HM}}$ is proportional to the difference in Height between the top of the barrier and the false vacuum, with the potential evaluated in the Jordan frame and \emph{including} the $\frac{1}{2}\xi\varphi^2R$ term as if it were part of the potential.

Stability during inflation requires that the probability of decay is sufficiently low that the expected number of separate causal regions in which a bubble was nucleated in our past light-cone, is fewer than 1. 
The survival of a single causal region that decayed to the false vacuum could potentially destabilize the universe if it were to then continue expanding after inflation ended.\footnote{Though it is unclear precisely what would happen to such a bubble that started expanding during inflation, only for inflation to end, as opposed to a bubble forming in flat space - here we will assume that it expands and envelops the whole spacetime.} If $N$ e-folds of inflation are visible, there are approximately $e^{3N}$ such causal regions,
and so the number that decayed during inflation is
\begin{equation}
n_{\rm{decayed}} = e^{3N}p(N,1),
\end{equation}
where $p(N,1)$ is the probability of a single causal region decaying after $N$ e-folds of inflation. Note that a bubble forming during inflation always expands to fill the causal region (1 Hubble volume) that it fills\cite{vacstab14}, but can expand no further because the expansion of space-time out-paces the bubble wall, that can only move at causal velocities. The probability per unit time that a bubble forms within a Hubble volume during inflation is given by
\begin{equation}
\gamma = V_{\rm{Hubble}} c H^4 e^{-B_{\rm{HM}}} = \frac{4\pi c}{3}He^{-B_{\rm{HM}}}.
\end{equation}
Thus, the probability of decaying between e-folds $N_1$ and $N_2$ is
\begin{equation}
p(N_2,N_1) =\frac{4\pi c}{3} \int_{t_1}^{t_2}\dd t H e^{-B_{\rm{HM}}} = k\int_{N_1}^{N_2} \frac{\dd N}{N} e^{-B_{\rm{HM}}}.
\end{equation}
Where $\frac{\dd N}{N} = H\dd t$, and $N$ is the number of e-folds. $k$ is an unknown $O(1)$ factor. For constant Hubble rate, which we will assume here for simplicity, this means
\begin{equation}
n_{\rm{decayed}} = k \log(N)\exp\left(3N - \frac{8\pi^2\Delta V_{\xi}(\varphi_{\rm{HM}})}{3H^4}\right).
\end{equation}
The condition $n_{\rm{decayed}} < 1$ then translates to
\begin{equation}
H < A\Delta V_{\xi}(\varphi_{\rm{HM}})^{\frac{1}{4}},
\label{eq:Hbound}
\end{equation}
where
\begin{equation}
A = \left[\frac{8\pi^2}{3(\log k + \log\log N + 3N)}\right]^{\frac{1}{4}} = 0.617\pm 0.004,
\end{equation}
for $N = 60$ e-folds and the uncertainty given by assuming $10^{-2} < k < 10^{2}$, illustrating the weak dependence on $k$. We apply this condition to the potential computed in this paper, using the solution of Eq. (\ref{eq:muvanish}) with smallest logs at the top of the barrier, for a range of $\xi$ and $H$ - the results are plotted in figure \ref{fig:stability_analysis}. We have also checked that this stability analysis is not affected by using a different scale choice, such as $\mu^2 = a\varphi_{\rm{cl}}^2 + bR$: this produced virtually identical results to figure \ref{fig:stability_analysis}.

The condition (\ref{eq:Hbound}) is of the same form but slightly strong than the bound used in \cite{vacstab7}, $\frac{8\pi^2V(\phi)}{3H^4} > 1$ which corresponds to  $A = (8\pi^2/3)^{1/4}\simeq 2.26$.  
As discussed in Refs.~\cite{vacstab14,vacstab13,East:2016anr,PhysRevD.97.025012}, the bound could be improved further by 
accounting for possibility for the field to flow back across the barrier due to evolution during inflation and by including the possible impact of CdL solutions. 

The effect of changing the top mass at constant $H$ is shown in figure \ref{fig:xibound}. Together these plots illustrate that even for Hubble rates somewhat below the instability scale (defined by $\lambda(\mu_{inst}) = 0$), negative $\xi$ can quickly destabilize the potential. Note also that because of the running of $\xi$, $\xi_{\rm{EW}} =0$ is qualitatively equivalent to having $\xi<0$, since the non-minimal coupling runs to negative values at higher energies, and if the optimal scale choice is $\mu^2 \sim R$ for small $\varphi$, this means $\xi < 0$ for the whole range of the potential for any non-zero Hubble rate.
\begin{figure}
	\centering
    \includegraphics[height=0.38\textheight]{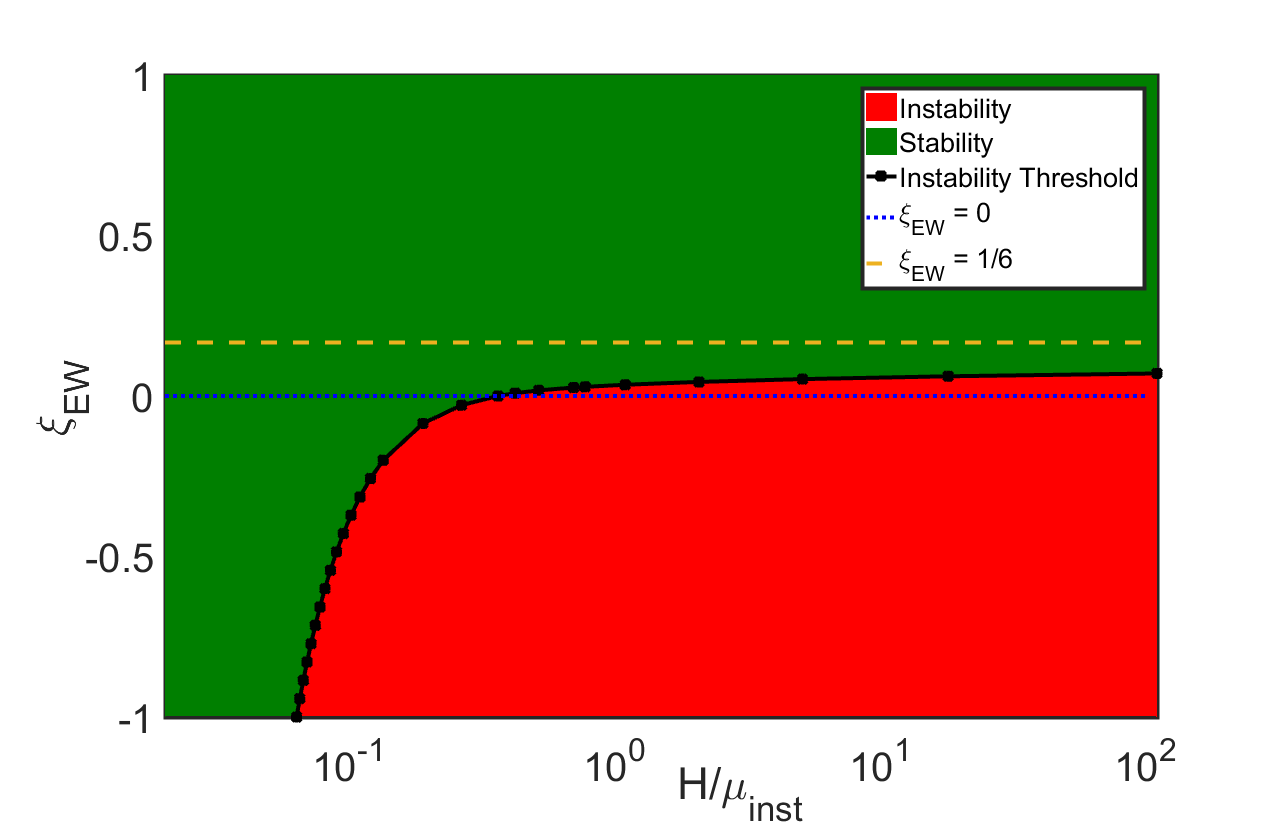}
	\caption{\label{fig:stability_analysis}Stability analysis for $M_t = 173.34\rm{ GeV}$, $M_h = 125.15\rm{ GeV}$. The red region has on average more than one bubble nucleation event within the observable universe during inflation, while the green region has less than one such event. $\mu_{\rm{inst}}$ is defined as the renormalization scale (in flat space) at which $\lambda(\mu_{\rm{inst}}) = 0$.}
\end{figure}
\begin{figure}
	\centering
    \includegraphics[height = 0.38\textheight]{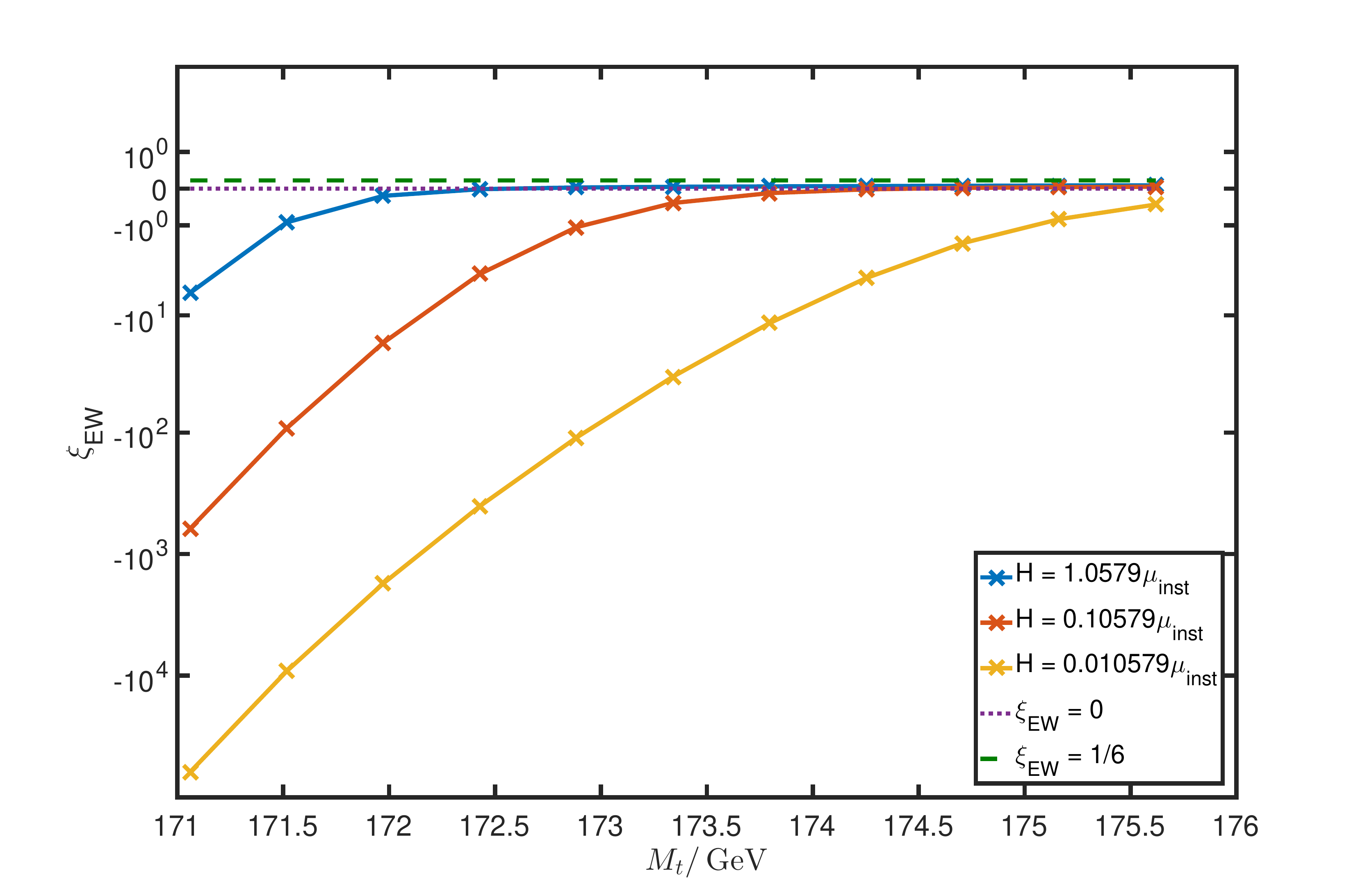}
	\caption{\label{fig:xibound}Plot of the boundary between stability and instability for different Hubble rates, and top masses. Note that $\mu_{\rm{inst}}$ is defined by $\lambda(\mu_{\rm{inst}}) = 0$ using $m_t = 173.34\rm{\,GeV}$ for comparison. Note that since the results in this paper use only 1-loop running, low values of $m_t$ can be unstable for values that would lead to an absolutely stable vacuum at 3-loops.}
\end{figure}
\FloatBarrier
\section{Conclusions}
In this work we derived the renormalization group improved effective potential in curved spacetime for the SM Higgs including the complete SM particle content to one loop order in perturbation theory. 
Our calculation included the UV limit of the loop corrections and thus contains the universal contribution that must be shared by all quantum states possessing the coinciding UV divergent behavior. 
We also presented the complete set of $\beta$-functions for the SM to one loop order, including all operators that are generated in curved spacetime. As an application we investigated the behavior of the SM Higgs in de Sitter space in the context of electroweak vacuum instability. 

Our use of the UV expansion means that the effective potential does not include infrared contributions, which can be large in the presence of light scalar fields. Locally the infrared contributions can always be absorbed into rescaling of background quantities and therefore do not affect our ultraviolet results. 
The global effects of these infrared modes can be studied by using the stochastic inflation approach~\cite{Starobinsky:1994bd} with the effective ultraviolet potential computed here as the input.  

Broadly speaking our results highlight two important, and often overlooked, aspects that arise whenever quantum fields are investigated in situations for which the curvature of spacetime is non-negligible: 
The first is that the renormalization group running sees the energy scale set by the curvature of the background, a mechanism which we called curvature induced running. For cosmologically interesting cases where the field is a light spectator with respect to the Hubble rate, the curvature can give the dominant contribution to the renormalization group running. The second aspect is the generation of operators invisible in flat space. In addition to the well-known non-minimal coupling there are 5 other operators (see (\ref{eq:treecurve})) generated via loops in curved spacetime. For the SM  in curved spacetime the $\beta$-functions imply the generation of all such operators resulting in important modifications, as is apparent from the results of section \ref{eq:gravb}. There is no compelling reason to assume similar contributions not to arise for theories beyond the SM, which can be studied by straightforward generalizations of our results.

The application to de Sitter space shows clearly the impact of making use of an effective potential calculated in curved spacetime. 
The standard procedure of optimizing the convergence of the loop expansion by an appropriate renormalization scale choice is made more complicated by the additional scale introduced by curvature. As we showed, even in the simple case of de Sitter space finding a physically motivated scale choice is non-trivial.
Using the effective potential, we demonstrated for the SM that negative values of the non-minimal coupling are tightly constrained from below by the requirement of vacuum stability during inflation. Importantly, this is true even for inflationary scales well below the scale of instability and hence for low top mass values for which the instability occurs above the maximal inflationary scale allowed by the non-detection of a primordial tensor spectrum. 

\acknowledgments
The authors thank Jos\'e Espinosa and Hardi Veerm\"ae for useful comments on the manuscript. 
TM and AR are supported by the STFC grant ST/P000762/1, and SS by the Imperial College President's PhD Scholarship. 

\bibliographystyle{JHEP}
\bibliography{vac_stab_long}

\end{document}